\documentclass{sig-alternate}
\usepackage{mathptmx} 

\newcommand{\ignore}[1]{}
\usepackage{fancyhdr}
\usepackage[normalem]{ulem}
\usepackage[hyphens]{url}
\usepackage{microtype}

\usepackage[pass]{geometry}
\usepackage{fancyhdr}
\usepackage[normalem]{ulem}
\usepackage[hyphens]{url}
\usepackage[table]{xcolor}
\usepackage{makecell}

\usepackage[font=small,skip=6pt,labelfont=bf,textfont=bf]{caption}
\usepackage{multirow}
\usepackage{mathptmx}
\usepackage{ragged2e}
\usepackage{dblfloatfix}
\usepackage{booktabs}
\usepackage{listings}
\usepackage{graphicx}
\usepackage{float}

\usepackage{bbding}
\usepackage{pifont}
\usepackage{wasysym}
\usepackage{amssymb}

\newcommand{\todo}[1]{}
\renewcommand{\todo}[1]{{\color{red} ({#1})}}

\usepackage{enumitem}
\setitemize{noitemsep,topsep=0pt,parsep=0pt,partopsep=0pt}
\setlist[itemize,1]{leftmargin=\dimexpr 26pt-.15in}

\newcommand{\chm}{\checkmark}

\usepackage{wrapfig}
\usepackage{balance}
\usepackage[bookmarks=true,breaklinks=true,letterpaper=true,colorlinks,linkcolor=black,citecolor=blue,urlcolor=black]{hyperref}

\renewcommand{\paragraph}[1]{\vspace{0.05in} \noindent\textbf{#1:}}

\fancypagestyle{firstpage}{
  \fancyhf{}

  \fancyfoot[C]{\thepage}
}  

\title{Exploiting Fine-Grain Ordered Parallelism \\  in Dense Matrix Algorithms\vspace{-0.5in}} 
\author{Jian Weng, Vidushi Dadu, Tony Nowatzki\\ \\ \normalsize University of California, Los Angeles\\ \normalsize \{jian.weng, vidushi.dadu, tjn\}@cs.ucla.edu}

\begin{document}

\maketitle
\thispagestyle{firstpage}
\pagestyle{plain}

\begin{abstract}
Dense linear algebra kernels 
are critical for wireless applications, and the oncoming proliferation of
5G only amplifies their importance.  Many such matrix algorithms are inductive,
and exhibit ample amounts of 
\emph{fine-grain ordered parallelism} -- when 
multiple computation flows with fine-grain producer/consumer dependences, 
and where the iteration domain is not easily tileable.  
Synchronization overheads make multi-core parallelism
ineffective, and the non-tileable iterations
  make the vector-VLIW approach less effective,
especially for the typically modest-sized matrices.

Because CPUs and DSPs lose order-of-magnitude performance/hardware utilization,
costly and inflexible ASICs are often employed in signal processing pipelines.
  A programmable accelerator with similar performance/power/area would be highly desirable.
We find that fine-grain ordered parallelism
can be exploited by supporting: 1. fine-grain stream-based communication/synchronization; 2.
inductive data-reuse and memory access patterns; 3. implicit
vector-masking for partial vectors; 4. hardware specialization of dataflow
criticality.

In this work, we propose, REVEL, as a next-generation DSP architecture.  
It supports the above features in its ISA and microarchitecture, and further
uses a novel vector-stream control paradigm to reduce control overheads.
Across a suite of linear algebra kernels,
REVEL outperforms equally-provisioned DSPs by 
4.6$\times$-37$\times$ in latency, and achieves a 
performance per mm$^2$ of 8.3$\times$.  It is only 2.2$\times$ higher
power to achieve the same performance as ideal ASICs,
at about 55\% of the combined area.
\end{abstract}


\section{Introduction} \label{sec:intro}

Dense linear algebra kernels, like matrix factorization, multiplication,
decomposition and FFT, have for decades been the computational workhorses
of signal processing across standards,
specifications, and device settings.  The oncoming proliferation of 5G
wireless is only further pushing the computational demands, 
both in performance and energy efficiency.  Driven by
needs of higher capacity and applications like augmented and virtual
reality~\cite{mobile-edge-whitepaper}, 
new standards will require signal processing at
more than an order-of-magnitude higher throughput and lower latency.

\begin{figure}
\begin{center}
\includegraphics[width=\linewidth]{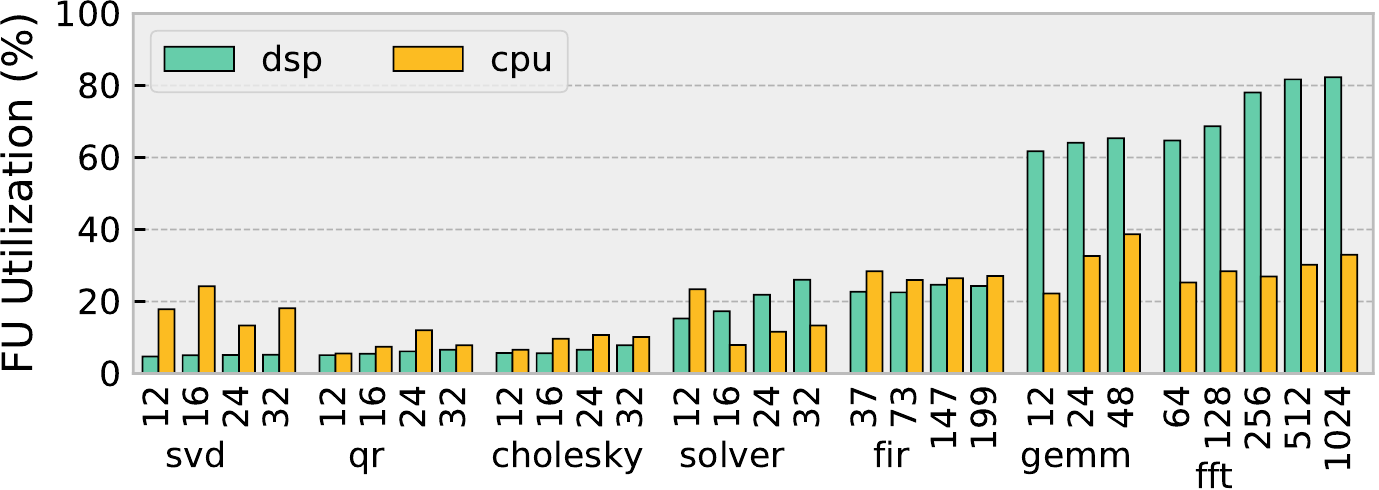}
\end{center}
\vspace{-0.198in}
\caption{Percent peak performance of CPU (Intel Xeon 4116) and DSP (TI C6678) on
DSP kernels}
\label{fig:cpu-dsp-util}
\vspace{-0.12in}
\end{figure}

Despite their ubiquity, many important dense matrix operations are far from trivial
to parallelize and compute at high hardware efficiency. 
As evidence, Figure~\ref{fig:cpu-dsp-util} shows the
hardware utilization (based on max. vector issue width),
of a modern CPU and DSP running common DSP algorithms from
native application suites (eg. MKL, and TI DSPLIB).  
For algorithms without fine-grain dependences (GEMM, FIR, and FFT),
a reasonable utilization is achieved, usually between 30-80\%.
However, for factorization/decomposition (SVD, QR, Cholesky, Solver),
the utilization is exceptionally poor, generally between 5\%-20\%.
Even this measure is generous as we only consider the maximum throughput
of a single core, yet there is enough raw parallelism to 
multithread.  Empirically, however, 
MKL and TI libraries do not even invoke multiple threads
at the commonly-small matrix sizes required, due to synchronization overheads.
CPUs and DSPs leave 
untapped factors of performance/hardware utilization.

\begin{figure}[b]
\begin{center}
\includegraphics[width=\linewidth]{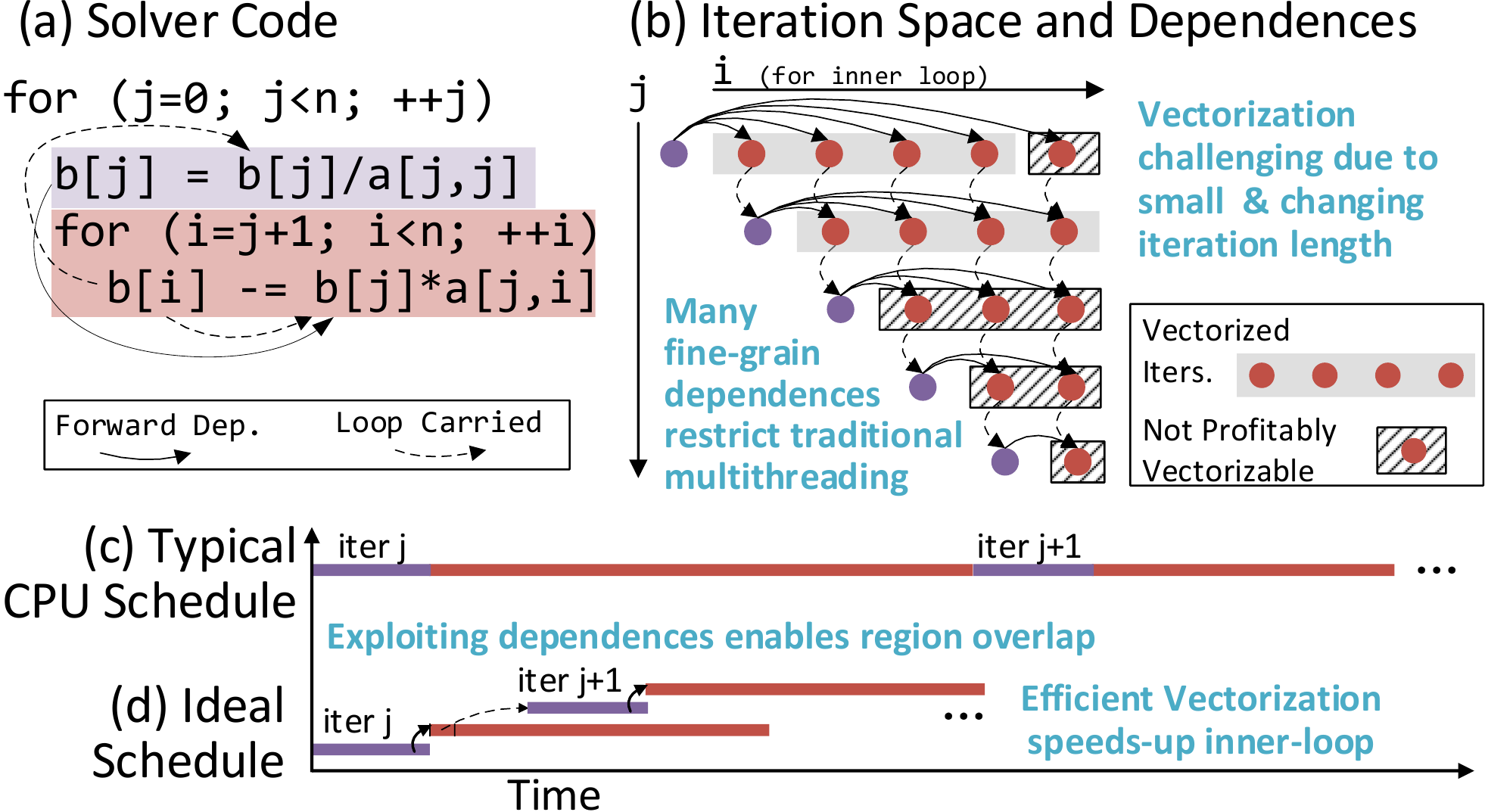}
\end{center}
\vspace{-0.17in}
  \caption{FGOP Example: Triangular Linear Solver} 
\label{fig:solver-fgop}
\vspace{-0.03in}
\end{figure}

The challenge and opportunity comes from the dominant form of parallelism in these
workloads, which we call \emph{fine-grain ordered parallelism} (FGOP).  FGOP consists 
of fine-grain producer/consumer relationships between otherwise parallel computations,
where the rate of production-to-consumption, the rate of data reuse, and the
memory access relation is an affine function of the induction variables.  
This results from the iterative and inductive nature of these algorithms, as they
operate on relatively small matrix sizes. 

To substantiate, consider the triangular solver in
Figure~\ref{fig:solver-fgop}(a).  Its iteration space diagram,
\ref{fig:solver-fgop}(b), reveals the many fine-grain dependences that make
profitable multithreading between regions impossible.  Furthermore, the
inner-loop trip count changes inductively, leading to many iterations that are
difficult to vectorize.  Nevertheless, an architecture can be designed to
exploit FGOP; the potential is shown in
Figure~\ref{fig:solver-fgop}(c,d).  If dependences between regions can be
enforced at a fine-grain with low overhead, then overlap between regions
becomes possible, increasing the parallelism.  If the inductive memory access
pattern (and its relationship to computation) can be expressed efficiently,
then vectorization can reduce the total time of the inner-loop region. 

ASICs can of-course be designed to exploit FGOP
-- hence why they are so commonly employed for these tasks.  Unfortunately,
they have significant drawbacks: design time and verification effort, extra
on-chip area, lack of flexibility, and lengthened time-to-market; these are
especially relevant for example domain of wireless, where standards are
continually changing and infrastructure costs are high.  A general and
programmable architecture exploiting FGOP could prove to be a worthy, if not
essential, replacement of traditional vector-VLIW DSP architectures.

\paragraph{Goals}  Our goals are twofold: 1. developing
abstractions and execution semantics to enable efficient 
expression of FGOP; and 2. applying these abstractions to create
an efficient programmable accelerator instance for DSP algorithms, 
capable of accelerating both FGOP \emph{and non-FGOP} workloads 
in this domain (eg. GEMM, filters). 

\paragraph{Approach}
Through an in-depth workload analysis, we find four essential 
architecture abstractions to express FGOP efficiently to hardware:
1. parallel dataflows with ordered 
communication channels.
2. to reduce control overhead, 
induction-variable dependent communication, memory access, and data-reuse.
3. for efficient vectorization, the implicit masking of non-vector-width-divisible iterations.  
4. for high hardware utilization, the specialization of compute hardware for 
critical versus non-critical dataflows.

While in principle the above abstractions can be added to a conventional ISA,
we choose a stream-dataflow ISA~\cite{stream-dataflow}, as its dataflow-based
computation and communication abstractions are simple to modify, and the resulting
accelerator can be performance/power competitive with DSPs. 
For the hardware implementation, we start with a simple design for one lane: a scratchpad connected 
to a coarse grain reconfigurable fabric (eg. similar
to some previous designs~\cite{lssd,plasticine,morphosys,dyser}).
We use multiple such ``lanes'' to scale up the design.

\begin{figure}
\begin{center}
\includegraphics[width=\linewidth]{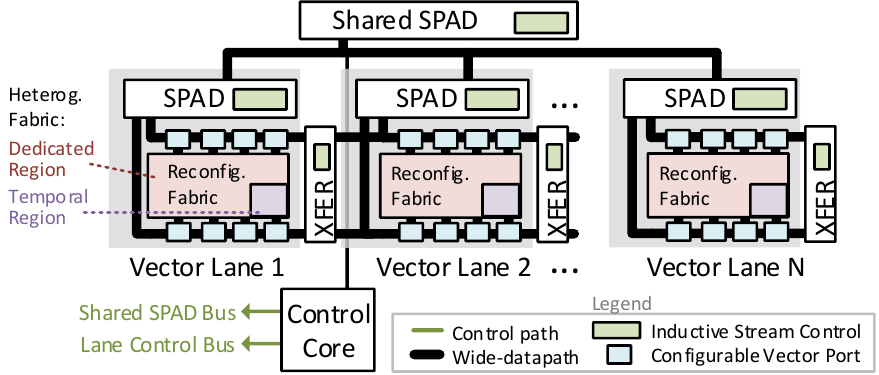}
\end{center}
\vspace{-0.18in}
\caption{REVEL Architecture Model}
\vspace{-0.06in}
\label{fig:arch-concept}
\end{figure}

Our accelerator, REVEL: the \underline{Re}configurable \underline{Ve}ctor \underline{L}ane architecture
(Figure~\ref{fig:arch-concept}), is constructed by adding support for each of
the FGOP-exploiting abstractions: 1. We allow multiple parallel dataflows
(similar to threads) which can communicate within/across lanes 
through FIFOs to support synchronization on fine-grain dependences.  
To simplify the ordering of commands, we centralize control into one 
control-core which coordinates all lanes.  
2. We provide the ability to express inductive memory access,
data-reuse and communication patterns by adding suitable state machines to FIFO
communication structures.  3. We implement implicit vector masking by
exploiting the relationship between computation-vector width and
communication-stream length.  4. For high computation utilization, we develop a
novel heterogeneous compute fabric,  where different regions are specialized
for critical and non-critical dataflows.



\paragraph{Our contributions are} 
\begin{itemize}
\item Identification and characterization of fine-grain ordered parallelism (FGOP)
as the main challenge for accelerating many dense linear algebra kernels.
\item Architecture and execution model for expressing FGOP naturally
to hardware.
\item Novel architecture features (vector-stream control, inductive access/reuse, implicit vector-masking, 
and heterog. fabric), enabling ASIC-like power/area/performance.
\end{itemize}

\paragraph{Results}  A single 1.25GHz REVEL unit
can outperform a 2.1GHz OOO core running highly-optimized MKL code on DSP
workloads by mean 9.6$\times$, with an area normalized speedup of 1308$\times$. 
Compared to a DSP, REVEL achieves between 4.6$\times$-37$\times$ lower
latency, with an area normalized speedup of 8.3$\times$. 
Compared to a set of ideal ASICs with equivalent performance, it is about 2.2$\times$ higher power 
and .55$\times$ the area.

\paragraph{Paper Organization}
We briefly motivate the kernels in Section~\ref{sec:background}, and  
analyze their challenges/potential in Section~\ref{sec:fgop}.  
The FGOP abstractions and ISA instance (REVEL) are in Sections~\ref{sec:exec-model}
and~\ref{sec:revel}.  Section~\ref{sec:uarch} and~\ref{sec:compiler} 
describe the microarchitecture and compiler.  Methodology and results are in
Sections~\ref{sec:methodology} and ~\ref{sec:eval}. We finally
cover related work and conclude.

\section{Why these workloads?} \label{sec:background}

\begin{figure}[b]
\begin{center}
\includegraphics[width=\linewidth]{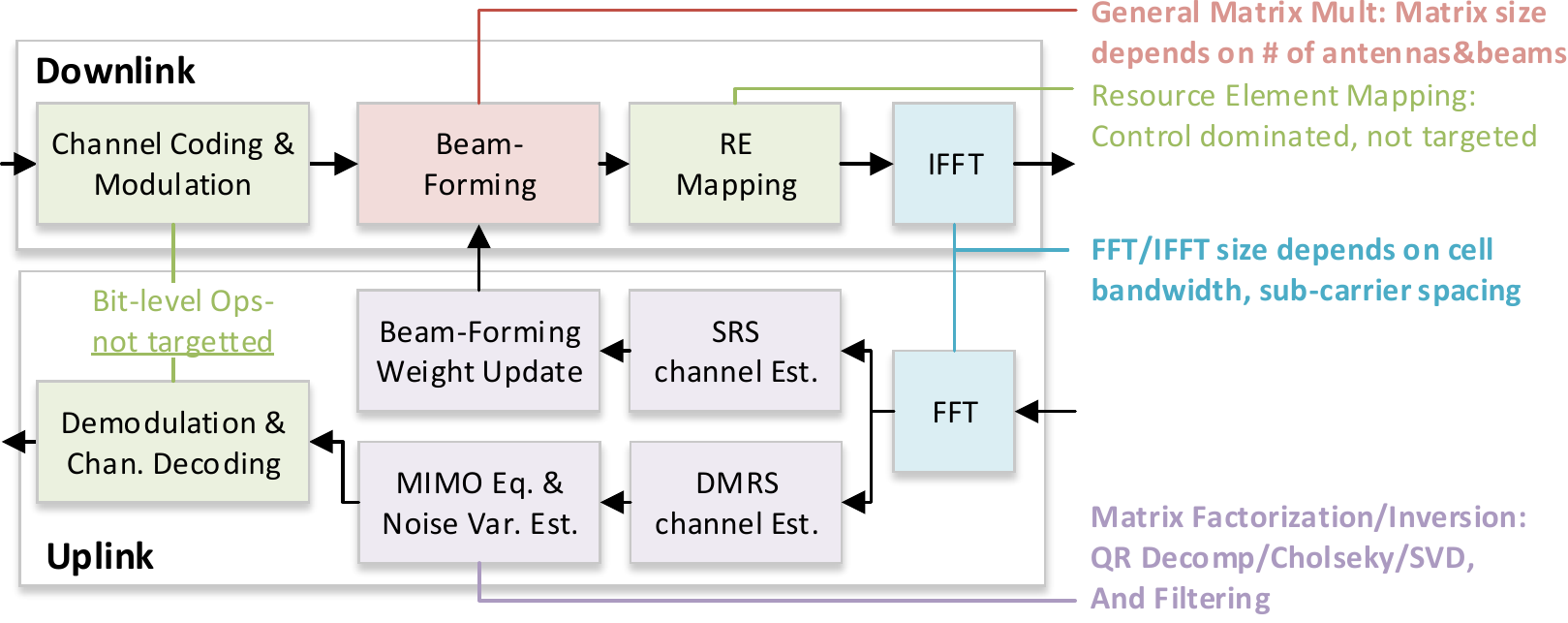}
\end{center}
\vspace{-0.13in}
\caption{Typical 4G/5G Transmitter/Receiver Pipeline}
\vspace{-0.04in}
\label{fig:5gpipes}
\end{figure}

We examine these DSP workloads as they represent a coherent and important set,
and because exploiting FGOP is critical for their performance.
To elaborate, Figure~\ref{fig:5gpipes} shows the stages of a
typical 4G/5G transmitter/receiver.

\paragraph{Kernels we do not target}
Channel coding and modulation involve mostly bit-level arithmetic.  RE mapping is a short resource
allocation phase which is not computation intense.

\paragraph{Kernels we target} The Beamforming stage involves mostly \textbf{GEMM},
coming from spatial signal filtering~\cite{beamforming}.
\textbf{Filters} and \textbf{FFT} of several varieties are also
common~\cite{polyphase,centro-symm,half-band}.

Challenging FGOP workloads are mostly within MIMO equalization and channel estimation.  These 
include \textbf{Singular Value Decomp.},
used for noise reduction~\cite{speech-svd},  \textbf{QR
Decomposition}, used for signal detection for 
(MIMO)~\cite{qr-dsp}, and \textbf{Cholesky decomposition} used in channel
estimation and equalization~\cite{cholesky-dsp,lmmse-wcdma}. 
\textbf{Solver} is an instructive example.

\paragraph{Why are matrices small?} The parameters often
depend on the number of antennas and beams (in the range of 
12-32 would be common)~\cite{wirelessdsp}.



\section{Fine-grain Ordered Parallelism} \label{sec:fgop}
Here we first define FGOP properties with
an example kernel, and explain why each is important either as a challenge or opportunity.
Then we characterize their prevalence in our workloads and
beyond. Finally, we perform a case study to answer why task-parallelism plus
vectorization is insufficient.

\subsection{Characteristic Properties of FGOP} 
\begin{figure}[tbp]
\includegraphics[width=0.99\linewidth]{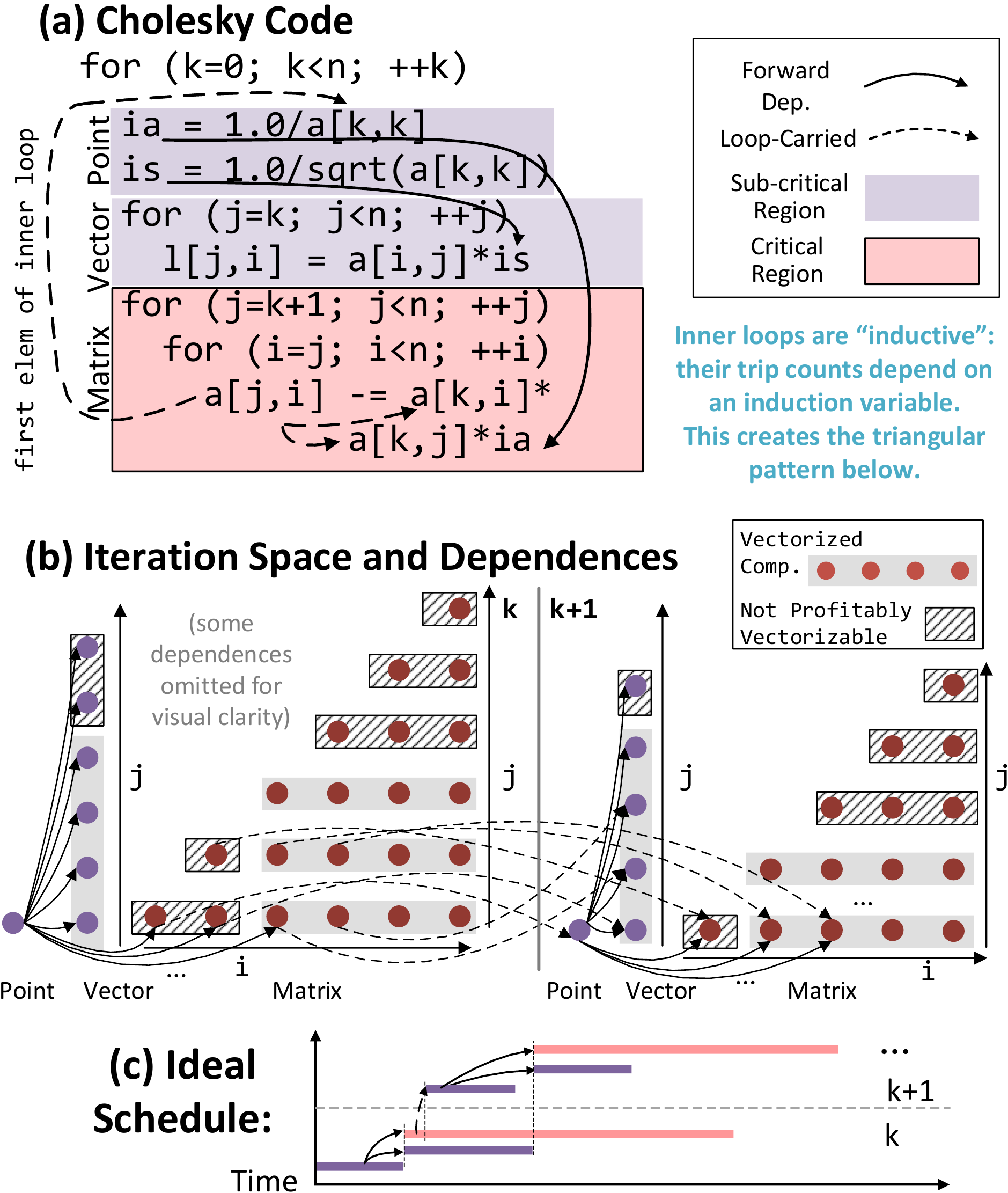}
\vspace{-0.05in}
\caption{FGOP Example: Cholesky}
\vspace{-0.09in}
\label{fig:fgop}
\end{figure}

We use
Cholesky decomposition as a running example in Figure~\ref{fig:fgop}.  Cholesky
contains a point, a vector, and a matrix computation region.  In general,
regions correspond to computations either across subsequent loops, or from
within the same imperfect loop but at different nesting levels.



\paragraph{Property 1: Parallel Flows with Fine-grain Dependences}  
A central characteristic of FGOP is the presence of fine-grain 
dependences between regions. In Cholesky, the vector and matrix region are
dependent on the point region (forward dep.), and the point region is dependent
on the first element in the matrix region (loop carried dep.). For a small
matrix, the granularity of these dependences is a few hundred instructions or less,
and even lower as the algorithm progresses.

\emph{Why is this important: the presence of these fine-grain dependences is the key
limiter to performance of multithreading the regions, due to high synchronization
overhead.}



\paragraph{Property 2: Ordered Dependences} 
Dependences are often strictly-ordered from the perspective of their
producing and consuming instructions.  
Figure~\ref{fig:fgop}(b) shows Cholesky's iteration space and dependences.  
In Cholesky, across multiple iterations of the outer
\texttt{k} loop, the point region is producing values (\texttt{inva} and \texttt{sqrt}) that are
consumed by the matrix region.  Similarly, the matrix region produces values consumed by
subsequent iterations of the point region.  
An example of a non-ordered dependence is when an array is consumed in the backwards order
of how it was produced.

\emph{Why is this important: the structure
of ordered dependences makes fine-grain synchronization of these dependences natural, and
therefore creates an opportunity to exploit efficiently in hardware.}

\paragraph{Property 3: Inductive Access/Data-Reuse} An inductive algorithm
iteratively builds on previous computations. In array codes, this often
manifests as induction-variable dependent trip-counts.  This is the case for
both of Cholesky's loops (but would not, for example, be true of a matrix
multiply).

This has implications for dependences, in that their reuse pattern 
(the rate of production to consumption) would be 
induction-variable dependent.  For example, how many times \texttt{inva} is consumed 
in the matrix region varies with \texttt{k} and \texttt{j}. 
Another example is that the matrix region only produces
a value for the next point region on the first iteration of the inner loop 
(so depends on \texttt{k}).

Also, inductive loops cause patterns of computation/memory
that are not easy to tile with vectorized loop iterations.  
Figure~\ref{fig:fgop}(b) also shows a vector-tiling pattern for Cholesky, with
many leftover iterations.   In a traditional vector architecture, these  
would require scalar iterations or masking/merging overhead.


\emph{Why is this important: inductive patterns cause overheads for coordination
of vector computation, as well as the wide interface to memory.}


\paragraph{Property 4: Region Imbalance}
Finally, regions often express imbalanced amounts of work.  In Cholesky, the
matrix region performs much more computation than the others, making it \emph{critical}
for performance.  In the Figure~\ref{fig:fgop}(a), we highlight
the critical region in red, and the sub-critical regions in purple. 
In DSP workloads, 
sub-critical regions often contain high-latency operations like \texttt{sqrt} and 
\texttt{div}.  

\emph{Why is this important: for a high hardware utilization, execution 
resources should be allocated appropriately to regions.  Furthermore, we
will show how it is possible to specialize the computation substrate for
criticality.}

\begin{figure}[b]
\includegraphics[width=0.99\linewidth]{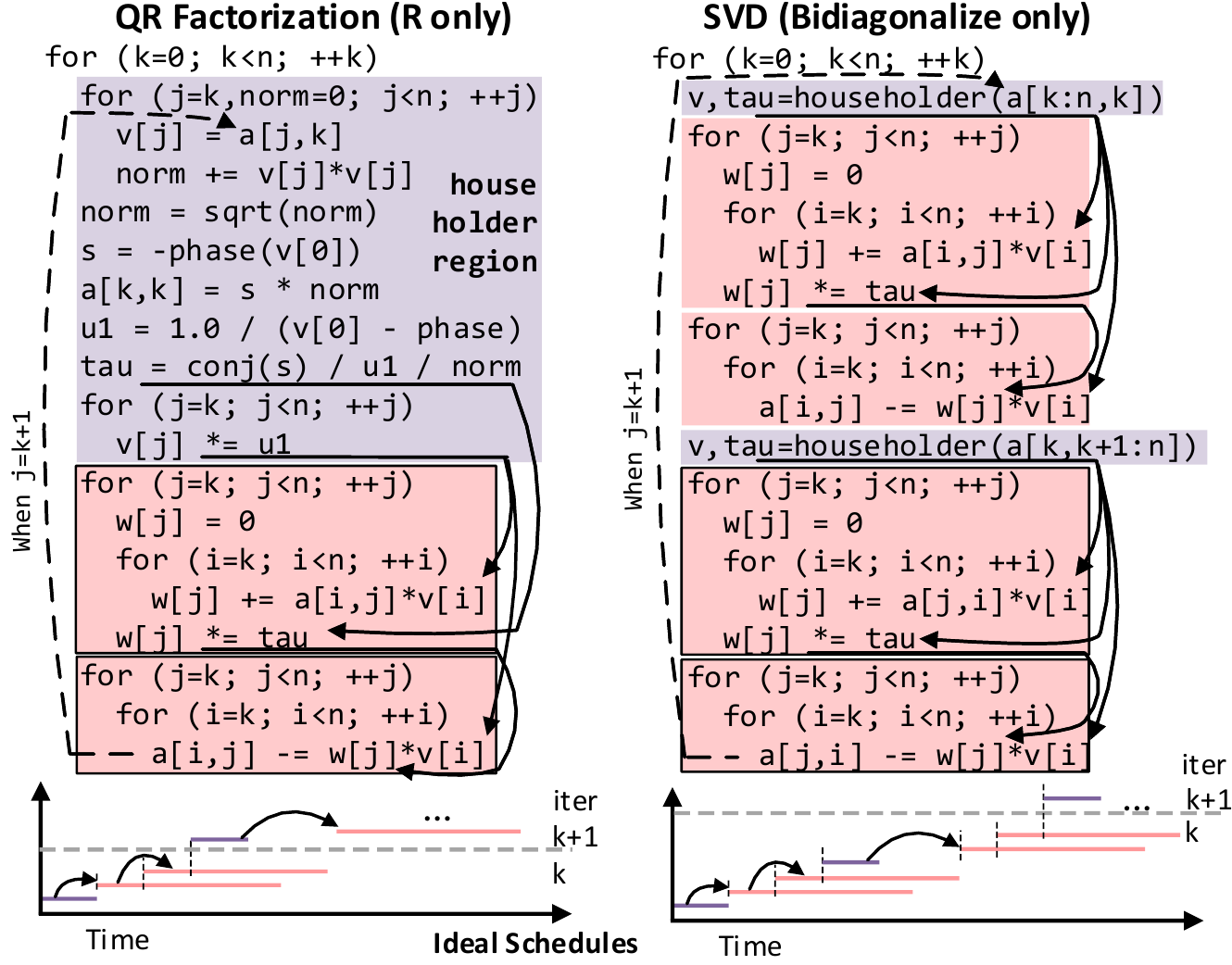}
\vspace{-0.05in}
\caption{FGOP Examples: QR and SVD}
\vspace{-0.05in}
\label{fig:other}
\end{figure}

\paragraph{Similar properties in QR, SVD} Figure~\ref{fig:other} shows that
both have fine-grain ordered deps. between
scalar/matrix regions (eg. \texttt{tau}) and between inner loops (eg. \texttt{w[j]}).
Inner loops are inductive and imbalanced compared to \texttt{householder} region.


\begin{figure}
\begin{center}
\includegraphics[width=\linewidth]{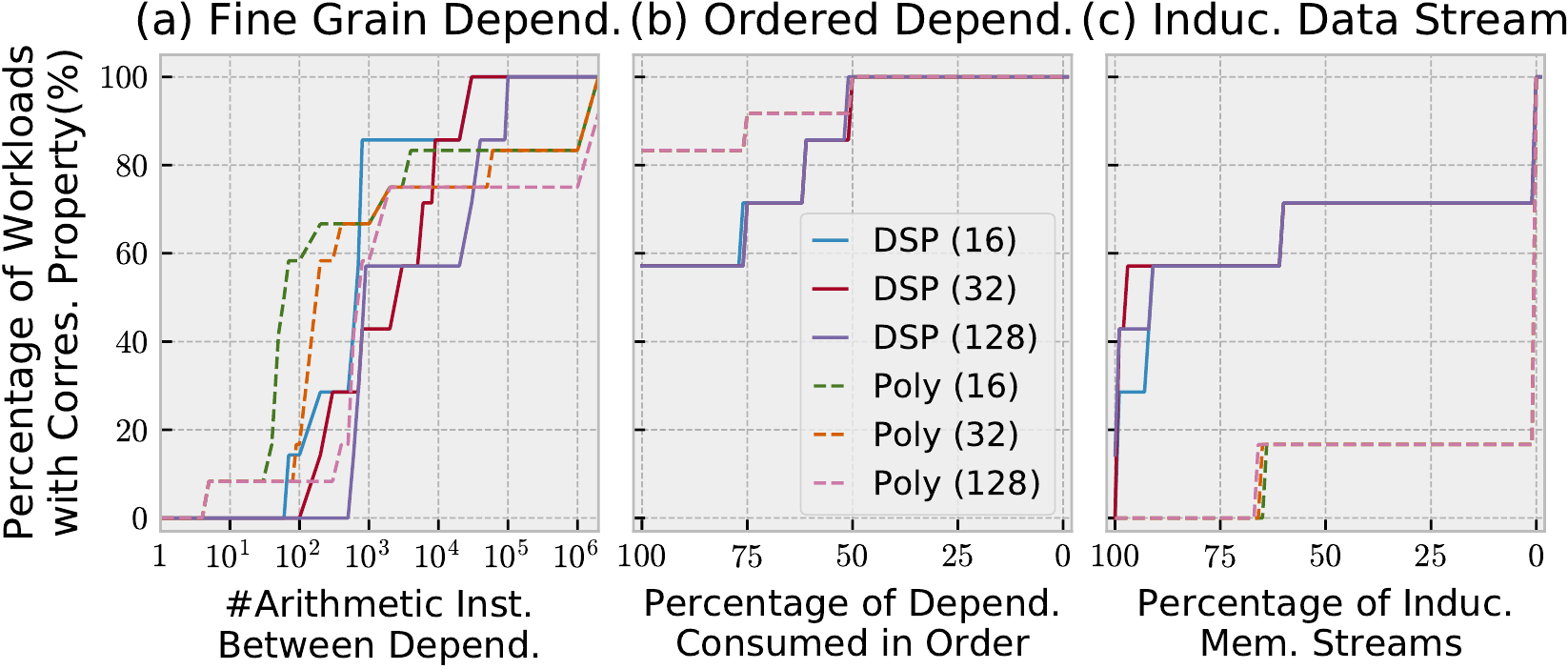}
\end{center}
  \vspace{-0.13in}
\caption{Prevalence of FGOP Properties.}
\label{fig:prevalence}
  \vspace{-0.06in}
\end{figure}

\paragraph{Prevalence of FGOP}  We characterize the prevalence of each FGOP
property in our 7 DSP workloads (Cholesky, QR, SVD, Solver, FFT, GEMM, Filter), as
well as a more general dense matrix benchmark suite,
PolyBench~\cite{polybench}.  We use LLVM~\cite{llvm} to
instrument programs to track dynamic memory dependences.
Figure~\ref{fig:prevalence} shows a cumulative density function (CDF) for each
property across three different matrix sizes (16,32,128)\footnote{FFT/Filter
does not process matrices, so we pick data size compared to the matrices.}.
\emph{In all plots, lines closer to the upper left indicates more FGOP.}

\begin{itemize}
  \item \textbf{Fine Granularity:} Figure~\ref{fig:prevalence}(a) characterizes the distance between inter-region
    dependences in terms of arithmetic instructions.  
    Most dependences (where the steepest part of the CDF curve is) are
    between about 75 to 1000 instructions, where larger data sizes are on the higher end.  Intuitively, this
    is a range where an out-of-order (OOO) core's instruction window
    begins to be insufficient, but still where shared-memory based synchronization also significantly 
    hurts performance (especially considering pipeline serialization during synchronization).
  \item \textbf{Ordered:}    Figure~\ref{fig:prevalence}(b) shows the prevalence of ordered dependences
    as a fraction of total dependences.  All workloads contain at least 50\% ordered dependences, and more
    than 80\% of DSP workloads are completely ordered; this is quite high and promising for later exploitation.
  \item \textbf{Inductive:}  DSP workloads have significant amounts of inductive access.  4/7 of the DSP workloads have
    more than 85\% inductive accesses.  PolyBench in general has much less inductive access, but still about 
    1/5 of those workloads are 60\% inductive. Nevertheless,
    the inductive property is critical for our DSP workloads.
  \item \textbf{Imbalanced:} 4/7 DSP workloads have imbalanced regions,
    while 50\% of PolyBench have imbalanced regions.
\end{itemize}

\emph{Overall, FGOP properties are common across dense matrix workloads, especially for
the relevant DSP workloads.}

\subsection{Why not task parallelism + vectorization?} \label{sec:challenges-existing}
We know from the data in Figure~\ref{fig:cpu-dsp-util} (page~\pageref{fig:cpu-dsp-util})
that exploiting FGOP is non-trivial for vector and VLIW cores. But why is this so, given
that DSPs are designed for these workloads?

The traditional method of parallelizing workloads with FGOP is through task
parallelism on a block of computations (eg. a set of iterations).  
Each dependence, or set of dependences, is simply a condition
under which to start a new task or synchronize existing tasks.  Intuitively, this works well 
when dependences are coarse grain (less overhead to start or synchronize), and when blocks are
perfectly tile the iteration space.  
As we discussed earlier, this is not true for the DSP workloads we study. 

\begin{figure}[t]
\begin{center}
\includegraphics[width=0.99\linewidth]{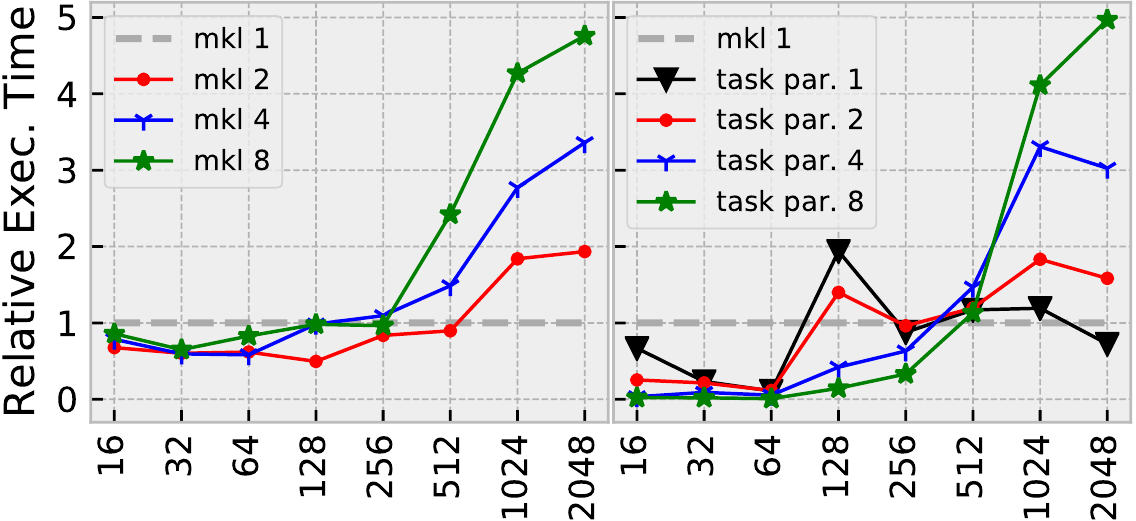}
\end{center}
\vspace{-0.15in}
  \caption{Case Study: Speedup of task-parallel Cholesky and MKL.} \label{fig:task-para}
\vspace{-0.02in}
\end{figure}

\paragraph{Case Study: Task-parallel Cholesky} 
For practical analysis, we analyzed an established Cholesky
kernel~{\cite{ompchol}} which uses blocked task-parallel
execution. 
Figure~{\ref{fig:task-para}} shows the task-parallel speedup over the single-threaded industry-standard MKL for
different matrix sizes (see
Sec~{\ref{sec:methodology}} for CPU and methodology).  
First, notice the its performance is similar to MKL 
for large matrices, which is possible because it calls underlying BLAS
routines (\texttt{dpotf2},\texttt{dtrsm},\texttt{dsyrk}) at a block-level.
We suspect MKL's implementation uses a similar approach, but does not
use task parallelism at small matrix sizes for performance reasons.

Considering the task-parallel code, the results indicate that exploiting
FGOP is only \emph{profitable} at all for larger matrices.  Using more threads simply
results in higher overhead of synchronization, far outweighing the benefits
of parallelization.  For the task-parallel version, speedups higher than
2$\times$ are only possible with matrices of 1024k size or larger, higher than we can make
use of in our domain.  Therefore, our goal is to create an architecture which can exploit FGOP better
at all matrix sizes \emph{and} enable speedup at small sizes.

\section{Abstractions to express FGOP} \label{sec:exec-model}

In this section we propose a set of architecture features
which can express fine-grain ordered parallelism efficiently to hardware.  
The description here is architecture-neutral, and we later develop an
architecture instance (an ISA).

\begin{figure}[b]
\includegraphics[width=\linewidth]{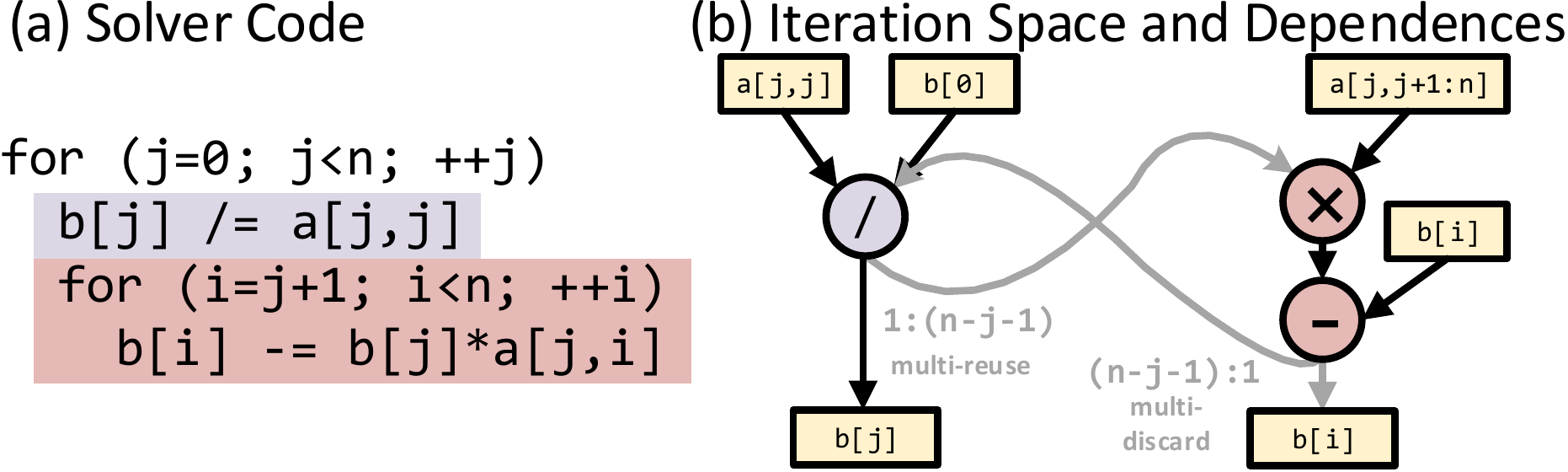}
  \vspace{-0.2in}
\captionof{figure}{Solver's Dataflows and Ordered Dependences}
  \label{fig:solver-dfg}
  \vspace{-0.05in}
\end{figure}

\paragraph{Feature 1: Concurrent Dataflows with Ordered Dependences}  
The essential abstraction is that of concurrent 
dataflows (threads) with the ability to express ordered dependences between regions.  
Ordered dependences are distinguished from typical instruction dependences in that
they have a non-uniform rate of \emph{production} to \emph{consumption}.  A consumption rate higher than
one indicates reuse of a value along a dependence.  
This may occur because data is reused multiple
times within a subsequent inner loop.
A production rate higher than one means that
several iterations occur before producing a value.
For example, this could be because data is being reduced (accumulated) for several iterations. 
Figure~\ref{fig:solver-dfg} shows the
\texttt{solver} kernel's dataflows, annotated with the memory access it performs within
each iteration of the outer \texttt{j} loop. Edges are labeled with their 
production:consumption rate, unless they are uniform (1:1).

\paragraph{Importance of Control Overhead}
One important consideration is the control-to-computation ratio.
Short-vector SIMD is one way to reduce control overhead;
one SIMD instruction expresses multiple operations over a fixed
number of data items.  A generalization  
(used in a variety of prior 
architectures~\cite{imagine,rsvp,stream-dataflow,q100,corampp,apmc,fpca})
is the concept of \emph{streams}, where a single control
command describes an entire pattern of operations.  The following features (2-4)
are related to the use of streams to reduce control overhead.

\begin{figure}
\begin{minipage}{0.48\linewidth}
\small
\begin{lstlisting}[mathescape=true]
for j=0 to $n_j$
  for i=0 to $n_i$
    array[$j * c_j + i * c_i$] 
\end{lstlisting}
\vspace{-0.07in}
  \textbf{(a) 2D Rectangular (RR)}
\end{minipage}
\begin{minipage}{0.48\linewidth}
\small
\begin{lstlisting}[mathescape=true]
for j=0 to $n_j$
  for i=0 to $n_i + j*s_{ji}$
    array[$j * c_j + i * c_i$] 
\end{lstlisting}
\vspace{-0.07in}
  \textbf{(b) 2D Inductive (RI)}
\end{minipage}
    \vspace{-0.04in}
  \captionof{figure}{Memory Address Stream Type Comparison} \hrule
    \label{fig:stream-types}
    \vspace{0.04in}
\includegraphics[width=\linewidth]{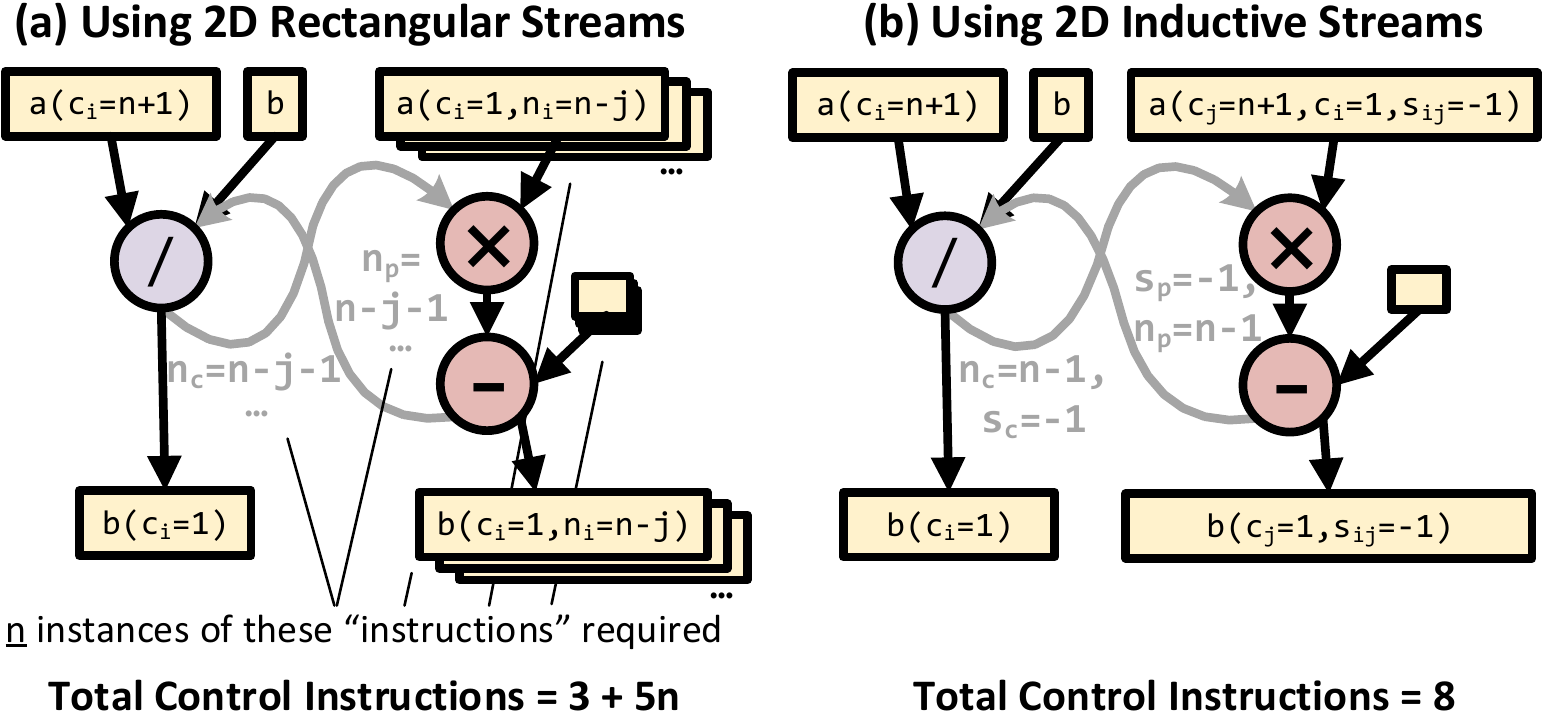}
  \captionof{figure}{Stream Specification using Different Types}     \hrule
  \label{fig:solver-streams}
\vspace{0.04in}
\includegraphics[width=\linewidth]{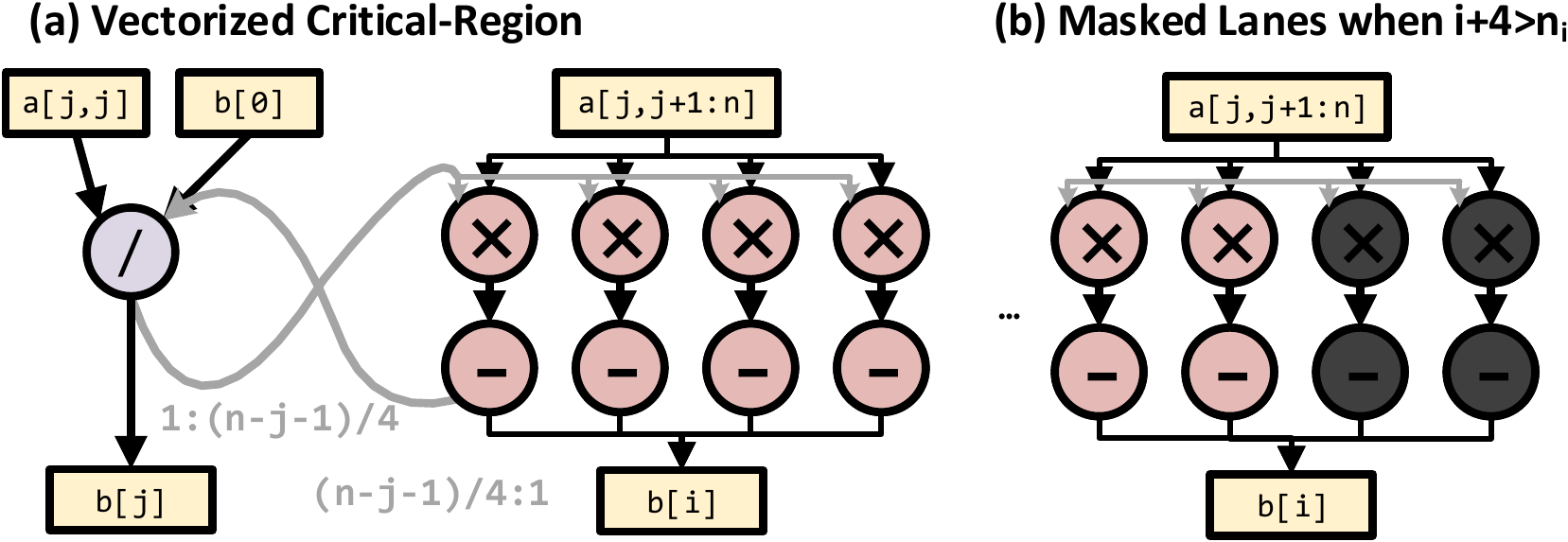}
  \vspace{-0.15in}
\captionof{figure}{Implicit Vector Masking}
  \label{fig:solver-vector}
  \vspace{-0.09in}
\end{figure}

\paragraph{Feature 2: Inductive Production:Consumption Rate}  
Data-reuse patterns may depend on induction variables, as seen in 
Figure~\ref{fig:solver-dfg}.  Here, the output of division is used multiple times
within the inner loop, but the number of times is reduced by one each iteration.  
In general we find that the pattern
changes only by small constant numbers.  We specify these as
two ``stretch'' parameters:  $s_{p}$ and $s_{c}$, 
the rate of change of production and consumption.
Figure~\ref{fig:solver-streams} contains an example encoding as a stream.
Including these parameters is not necessary for correct enforcement of dependences,
because multiple lower-dimension streams can be generated.
However, the number of instructions increases by an order of magnitude (as shown in Figure~\ref{fig:solver-streams}).

\paragraph{Feature 3: Inductive Memory Streams}  
Similarly, all prior architectures that we are aware of use \emph{rectangular} memory access streams: ie. their
iteration domains (without loss of generality) begin at $\vec{0}$ and end at a constant
$n$ in each dimension (ie. a \emph{trip count}),  and their address functions are linear
functions of $\vec{I}$.  If we let $c_i$, $c_j$ etc. be the multipliers of $\vec{I}$
in the address function, rectangular streams can then be depicted intuitively as a loop nest -- see
Figure~{\ref{fig:stream-types}(a)}. 

\emph{Inductive} streams are more general; their 
iteration domains may be bounded polyhedra instead of strictly rectangular.  
Trip counts become a linear function of lexicographically previous iterators.
We encode using
\emph{stretch} multipliers $s_{ji}$, representing the multipliers of iterator $j$ in the trip
count for dimension $i$.  Figure~{\ref{fig:stream-types}(b)} shows a 2D inductive stream pattern
as a loop nest.
Figure~{\ref{fig:solver-streams}} shows
how to specify the accesses in \texttt{solver} using either rectangular or
inductive streams.
Again, inductive access streams require O(n) fewer control insts.



Later evaluation uses a simple notation to describe capabilities:
Letter ``R'' denotes a rectangular dimension, and ``I'' denotes inductive dimension,
so ``RI'' would be a 2D capability with induction in second dimension.

\paragraph{Feature 4: Stream-based Implicit Vector Masking}
There are two issues with vectorization of FGOP. The first 
is that the reuse rate may become 
fractional, as it may be divided by the vector width (see example in 
Figure~\ref{fig:solver-vector}(a)).  Therefore, we need $s_{ij}$ to be able to represent
fractional numbers.  Second is 
the problem of non-vector-width divisible iterations.  To address this, we make it
implicit that the datapath for the remaining iterations becomes masked or predicated.
This can be enforced by dynamically checking the stream iterator for the case when
the inner-loop iterator $i$ is greater
than the current length $n_i$ (see Figure~\ref{fig:solver-vector}(b)).

\paragraph{Feature 5: Specification of Dataflow Criticality}  
Certain regions may be more or less computationally critical than others, as
they perform more or less work.  In our example in Figure~\ref{fig:solver-dfg},
the ``divide'' dataflow happens $n/2$ fewer times.  
In practice, non-critical dataflows should be allocated
shared resources, while critical dataflows should be vectorized. 
We will later demonstrate the effectiveness of hardware specialization for criticality.


\begin{figure*}[t]
\includegraphics[width=\linewidth]{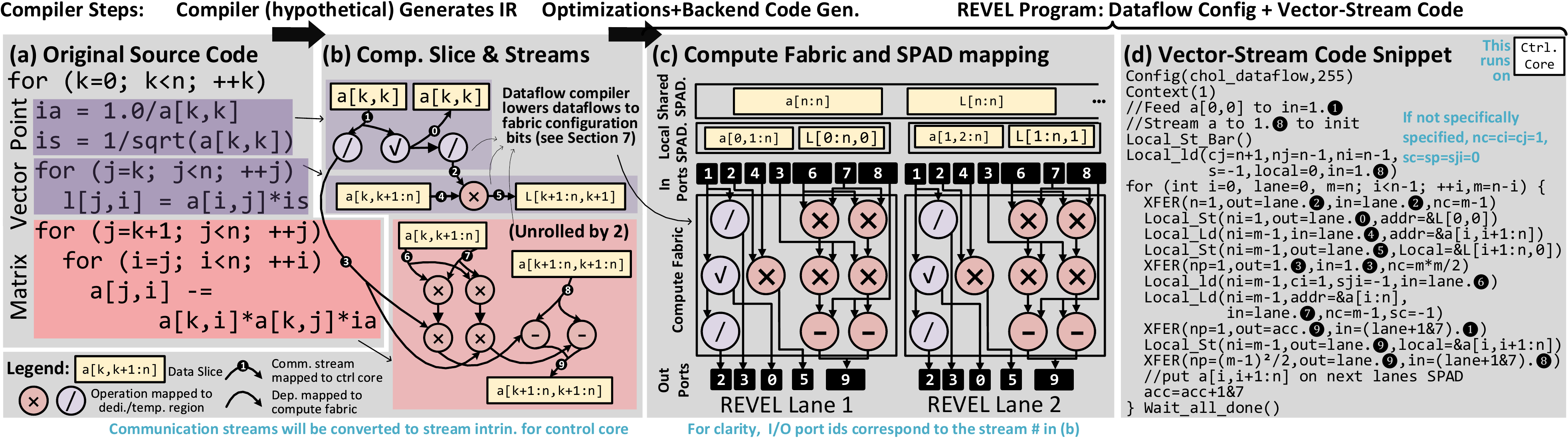}
\vspace{-0.16in}
\caption{Explaining REVEL abstractions using Cholesky as an example.}
\vspace{-0.02in}
\label{fig:rv-cholesky}
\end{figure*}

\section{REVEL: An FGOP-enabled ISA} \label{sec:revel}

Using the principles from the previous section, we construct an efficient
and scalable ISA and execution model (REVEL ISA) to 
exploit FGOP and traditional parallelism in dense matrix algorithms. 
REVEL is an instance of a stream-dataflow ISA~\cite{stream-dataflow}, which we
chose because it is straightforward to enhance for FGOP, and it enables an
efficient programmable accelerator. Section~\ref{sec:general} discusses enhancing 
other architectures (OOO cores and Plasticine~\cite{plasticine}).
In this section, we discuss the control model, how the architecture incorporates FGOP features, and
then its specific ISA instantiation.

\paragraph{Background on stream-dataflow} Stream-dataflow ISAs express
computation as a dataflow graph, where its inputs and outputs are 
named ports.  Communication is performed using streams, where the endpoints
of streams are either the dataflow-graph ports or memory.  A VonNeumann program
embeds all stream commands, and streams with the same port number are guaranteed
to be executed in program order.  Memory requests can be ordered by explicit 
barriers.

\paragraph{ISA Enhancements to support FGOP}
\begin{itemize}
\item \textbf{Ordered Dependences between Dataflows:} Computation is expressed
as \emph{multiple} independently-triggered dataflow graphs, where streams describe
their communication and re-use pattern.
\item \textbf{Inductive Dependence/Access:} Stretch 
parameters (s$_{p}$,s$_{c}$) added to relevant streams.
\item \textbf{Vector Masking:} Non-divisible iteration lengths causes
predication of the corresponding dataflow.
\item \textbf{Execution Rate:} Implementation
  is closely related to hardware, so we describe
    separately (Section~\ref{sec:het-fab}).
\end{itemize}

\paragraph{Example: Cholesky}  
Figure~\ref{fig:rv-cholesky} demonstrates REVEL's abstractions by
showing the transformation from source (a) to the abstract dataflow IR (b), 
and finally to dataflow configuration and stream-code running on the control core (c,d).

\paragraph{Enabling Scalability with Lanes} To enable scalability at low
overhead, we chose to add multiple lanes of execution. Each lane is
independent, in that it can concurrently execute multiple dataflows, each
potentially communicating using inductive streams or through a global memory.
Also, since each lane can be programmed separately, the architecture is
flexible in terms of what computations are being parallelized.

\paragraph{Vector-stream control}
There are two challenges with using
multiple lanes: 1. Each lane needs coordination (control overhead), 
and 2. The dataflow-dependence streams between lanes must somehow be ordered. 

Our solution is the vector-stream control model.  Here, a single
VonNeumann control program coordinates the execution of all lanes.
Control commands are sent to
all relevant lanes, specified by a bitmask (eg. load array from address 0 of local memory to dataflow
1).  In addition, a lane's index can be used to offset the address of a
command, so a single command can allow each lane to read a different portion of
an array.  This is unique and more powerful than the control amortization
offered by either vectorization or streaming alone, as it amortizes both in ``space''
across lanes, and in time through streaming commands.  
It is inspired by Vector-threading~\cite{scalevt,mavenvt,vlt} 
but with a stream-based ISA.

In the example, Figure~\ref{fig:rv-cholesky}, we map all three dataflows
(scalar, vector, matrix) to one lane to share its resources, and parallelize
the outer $k$ loop across lanes. 




%
%

\begin{table}[t]
\setlength{\tabcolsep}{0.06in}
\footnotesize
\centering
\begin{tabular}{@{}lllll@{}}
\toprule
\multicolumn{1}{r}{} & Pattern Params                                       & Source Params                         & Dest. Params                                                                         &                                  \\ \midrule
Shared\_ld           & \multirow{2}{*}{$c_i$, $c_j$, $n_j$, $n_i$}          & shared\_addr                          & \multicolumn{1}{l|}{local\_addr}                                                     & \parbox[t]{2mm}{\multirow{8}{*}{\rotatebox[origin=c]{90}{~Lane Bitmask (ALL)~~~ }}} \\
Shared\_st           &                                                      & local\_addr                           & \multicolumn{1}{l|}{shared\_addr}                                                    &                                  \\ \cmidrule(r){1-4}
Local\_st            & \multirow{2}{*}{$c_i$, $c_j$ $n_j$, $n_i$, $s_{ji}$} & out\_port                             & \multicolumn{1}{l|}{local\_addr}                                                     &                                  \\ \cmidrule(lr){4-4}
Local\_ld            &                                                      & \multicolumn{1}{l|}{local\_addr}      & \multicolumn{1}{l|}{}                                                                &                                  \\ \cmidrule(r){1-3}
Const                & ~~~~~$n_j$, $n_i$, $s_{ji}$                    & \multicolumn{1}{l|}{val$_1$, val$_2$} & \multicolumn{1}{l|}{\begin{tabular}[c]{@{}l@{}}in\_port,\\$n_c$, $s_c$\end{tabular}}                                                                &                                  \\ 
  XFER                 & ~~~~~$n_p$, $s_p$                                    & \multicolumn{1}{l|}{out\_port}        & \multicolumn{1}{l|}{} &                                  \\ \cmidrule(r){1-4}
Configure            &                                                & local\_addr                           & \multicolumn{1}{l|}{}                                                                &                                  \\
\multicolumn{3}{l}{\hspace{-.07in}Barrier Ld/St \& Wait}                                                                      & \multicolumn{1}{l|}{}                                                                &                                  \\ \bottomrule
\end{tabular}
\vspace{-0.05in}
\caption{REVEL's Vector-Stream Control Commands}
\vspace{-0.06in}
\label{tab:commands}
\end{table}

\paragraph{REVEL Commands}
Table~\ref{tab:commands} contains the set of commands within the
VonNeumann control program for stream coordination, including their
pattern parameters, source, and destination.  \texttt{Shared\_Ld/St} are for
transfers between local and shared memory.  \texttt{Local\_Ld/St} transfer between the
local memory and the dataflow graph.  XFER specifies inter-dataflow communication
streams to support fine-grain dependences.
\texttt{Const} can stream a pattern of val$_1$,val$_2$, eg. (\underline{0,0,0},1,\underline{0,0},1,\underline{0},1), which is
useful for inductive control-flow within the dataflow graph.
The \texttt{Barrier\_Ld/St} command prevents concurrent scratchpad memory access, 
and \texttt{Wait} delays until a lane is no longer active.
These are used for flexible double buffering.
 All commands take a lane bitmask as a parameter, to implement vector-stream control.

\section{REVEL Microarchitecture} \label{sec:uarch}

We describe REVEL's microarchitecture by first
giving a broad overview, then explaining the key innovations that enable
efficient exploitation of FGOP.  We discuss the heterogeneous compute fabric 
in detail, as it is a key novel component of the
design, enabling low overhead execution of unbalanced FGOP regions.

\subsection{Hardware Overview} 
\begin{figure}
\includegraphics[width=\linewidth]{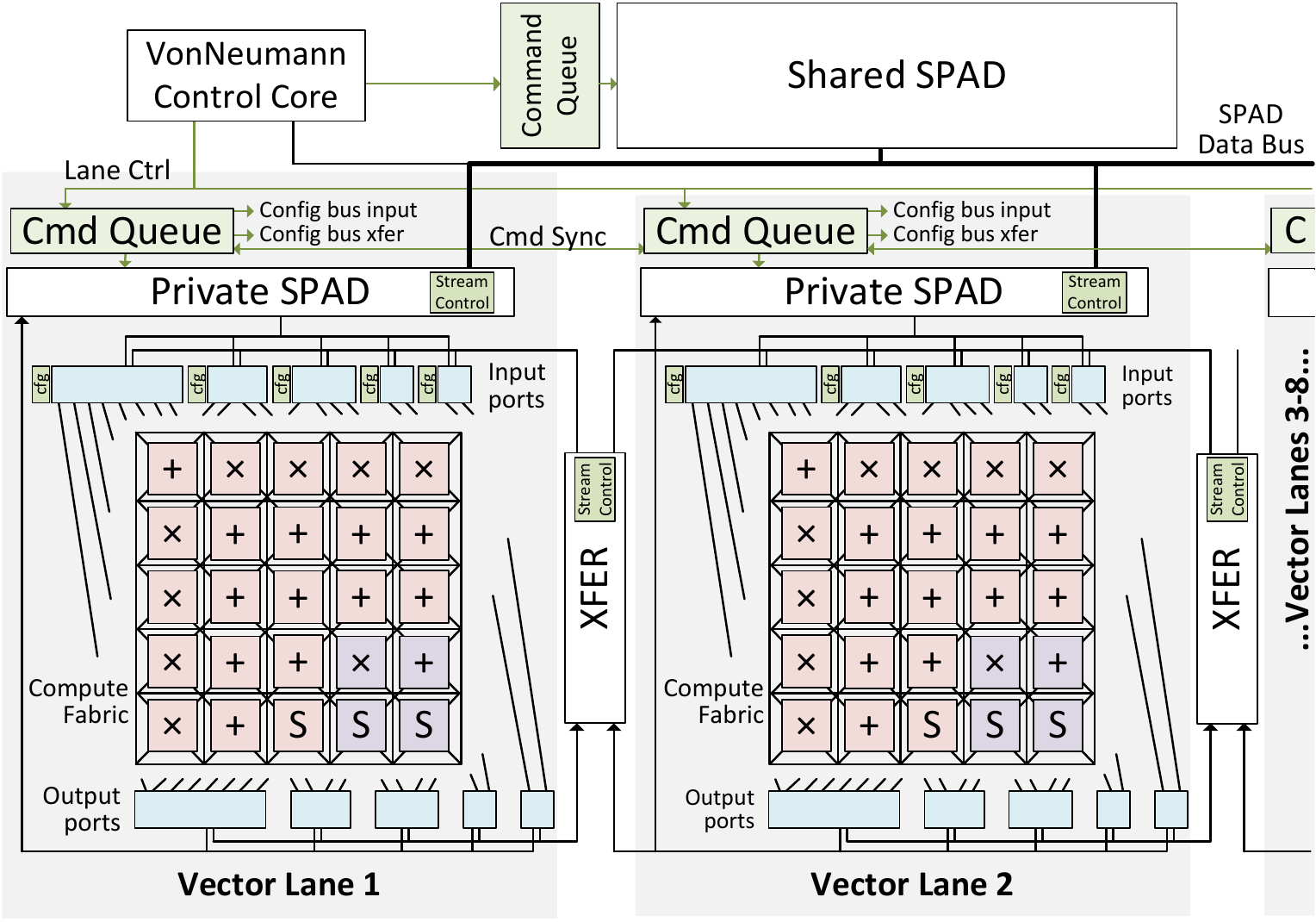}
  \vspace{-0.21in}
  \captionof{figure}{REVEL Microarchitecture}
\label{fig:revel}
  \vspace{-0.11in}
\end{figure}

The REVEL processor (Figure~\ref{fig:revel}) is
composed of a number of lanes, a low power VonNeumann control core, and a
shared scratchpad.  The control core can issue vector-stream commands to
each lane.  Each lane manages its stream and memory dependences,
data access requests and computation firing.  Dataflows on the same or separate
lanes can communicate data through the XFER unit or shared scratchpad.

\paragraph{High-level Operation}  REVEL's high-level 
execution flow is as follows:
First, the core will issue a \texttt{config} command, and 
configuration data is broadcast to each relevant lanes' 
compute fabric and its ports.
Asynchronously, the control core will begin to compute the parameters of any
stream commands.  When a command is ready, it will
be broadcast relevant lanes' command queues.  
Commands are then issued to either the private or  
shared scratchpad, provided the resources they depend on are free
(eg. input or output compute-fabric ports).
Streams execute locally until completion, until which point they notify 
the command queue
that they are free.  Independent dataflows may be mapped onto the compute
fabric, where they are executed in pipelined fashion once enough data has 
arrived for an instance of the computation. 
Once the control core has completed issuing vector-stream commands, it will
issue a \texttt{Wait} command.  This blocks the control program
until all relevant lanes are no
longer active, which is determined by the completion of all streams.

\paragraph{The responsibilities of each component are}

\begin{itemize}
\item \textbf{Command Queue} is the lane's resource manager, 
and is responsible for maintaining data ordering.  It 
maintains a queue of commands from the control core, and issues them
to the scratchpad or XFER unit if no barrier commands or port dependences prevent that.
A scoreboard tracks ports in-use.

\item \textbf{Stream Control}   maintains the set of concurrent
streams, where each stream tracks the state of its iterators
($i$,$j$) and length $n_i$, (to support inductive access).  It can
generate addresses for one stream per cycle, along with a mask for any unused
words of the scratchpad line.  Streams are prioritized by minimum ``cycles-to-stall,''
which is the number of cycles before the corresponding port will run out of
data (data-in-fifo / port-width). 

\item \textbf{Input/Output Ports}
contain a set of FIFOs for
holding intermediate results while waiting for (or produced from) the
compute fabric.  Input ports
can receive data either from the scratchpad bus, or from the XFER unit
if receiving data from a neighboring dataflow.  Each port attaches to a unique
location within the grid, so it is the compiler's responsibility to choose
optimal ports.

\item \textbf{Compute Fabric}  
monitors the data ready in each input port
FIFO to determine which dataflows can begin.  Multiple 
can be fired in a single cycle.
This heterogeneous fabric is divided into
regions which specialize for either critical or non-critical computations
    (Section~\ref{sec:het-fab}).

\item \textbf{XFER Unit}  is responsible for arbitrating the
bus from output ports back to the local or remote input ports, which enables
fine-grain dependences between dataflows, both within and across lanes.

\end{itemize}

\subsection{Supporting FGOP Abstractions}
Here we describe the essential hardware mechanisms for supporting
FGOP within REVEL.  Details on the heterogeneous compute
are discussed subsequently.

\paragraph{Concurrent Dataflows with Ordered Dependences}
To support multiple dataflows with different firing conditions, 
the data present in each dataflow's ports are tracked separately by the data-firing
logic, which can manage up to four independently-firing dataflows.  The association
between ports and dataflows is determined at configuration time.

One other challenge is maintaining data-ordering when there are fine-grain
dependences between lanes -- ie. a \emph{source} lane should not transmit until
all prior data items (in program order) for the \emph{destination} lane's port
have arrived.  This is accomplished by sending the destination lane a placeholder stream.
The destination's command queue
informs the source's when the placeholder is issued for the destination port, and the source's
command queue informs the destination's when the placeholder can be removed.

\paragraph{Inductive Memory Access} To support inductive re-use streams, the
scratchpad controller maintains the current iterator values \emph{and}
the current stream length.  When $n_i$ addresses are complete, the length is
adjusted by the stretch $s_{ij}$.  Note that $s_{ij}$ is a fixed-point number
to support vectorization with induction patterns.  

\begin{table*}[b!]
\setlength{\tabcolsep}{0.06in}
\small
\centering
\def\hyph{-\penalty0\hskip0pt\relax}
\footnotesize
  \begin{tabular}{@{}>{\RaggedRight}p{0.21in}>{\RaggedRight}p{0.2in}|>{\RaggedRight}p{1.1in}|>{\RaggedRight}p{0.9in}|>{\RaggedRight}p{1.1in}|>{\RaggedRight}p{1.6in}|>{\RaggedRight}p{1.0in}@{}}
\toprule
                                                                    &     & Ordered Dep.                                                                   & Inductive Dep.                           & Inductive Mem.                                       & Implicit Vector Mask                                                                           & Crit. Specialization                                          \\ \midrule
    \multirow{3}{*}{\shortstack{OOO \\ Core}}                                           & S/W & Thread-communicat-ion-aware OS sched &  (see below)                                                & Streaming memory command interface                        & Add FIFO interf. b/t streams \& vector instrs.                   & \multirow{3}{*}{\shortstack{Not applicable, no \\ explicit-dataflow \\ substrate.}} \\
                                                                    & ISA & Stream-based producer/consumer instrs.                                           & \multicolumn{2}{l|}{Add induction parameters to stream instrs.}                                                                 & Implicit mask register indicating predicated lanes. &                                                                     \\
                                                                    & H/W & Commun.-FIFOs b/t neighbor cores                                    & Add inductive control to FIFOs                   & Add streaming memory request engine  & Vector store instruction ignores masked lanes.                                                   &                                                                     \\ \cmidrule(lr){2-7}
    \multirow{3}{*}{\shortstack{Plast-\\icine\\\cite{plasticine}}} & S/W & \multirow{3}{*}{Already Supported}                                               & \multicolumn{2}{l|}{Add inductive param for map\&fold patterns}            & None                                                                                         & None                                                                \\
                                                                    & ISA &                                                                                       & \multicolumn{2}{l|}{Update stream-control and addr. gen. interf.}                              & Update stream instr. semantics                                                      & Temporal fabric ISA          \\
                                                                    & H/W &                                                                                       & Add induction to stream controller & Add induction to addr. gen.               & Implement predication within SIMD Lanes                                                      & Make some PCU's temporally shared                                \\ \bottomrule
\end{tabular}
  \vspace{-0.05in}
  \caption{Adding FGOP Abstractions to Existing Architectures \textnormal{S/W: Software, H/W: Hardware}}
  \label{tab:other-arch}
\end{table*}

\paragraph{Inductive Data Reuse}
While a stream without reuse would perform the usual 
destructive FIFO read on every cycle, a stream with reuse will
only pop the data from the port at a longer interval.  When the stream 
is issued from the command queue to the stream control unit, the reuse
configuration ($n_r$ and $s_r$) is sent to the port (maintained similarly
to params for memory access).
Besides enabling fine-grain dependences with inductive changes in re-use length,
another benefit of the reuse within the configurable port
is a large reduction in scratchpad bandwidth.

\paragraph{Implicit Vector Masking}  As a stream is executing, the stream
control unit compares the remaining iterations with the vector length of the
destination port.  If the iterations left is non-zero and less than the vector
length, the stream control unit sends the data padded with zeroes for the
unused lanes, along with additional meta-information which indicates that the
those lanes should be predicated off.  This information is buffered in a
predication FIFO which tracks the data FIFO.

\subsection{Heterogeneous Compute Fabric} \label{sec:het-fab}

Attaining high utilization in FGOP workloads requires
balancing execution resources between critical and non-critical dataflows.
This is especially challenging given that they prefer
quite-different fabric microarchitectures.  
\begin{figure}
\centering
\includegraphics[width=0.8\linewidth]{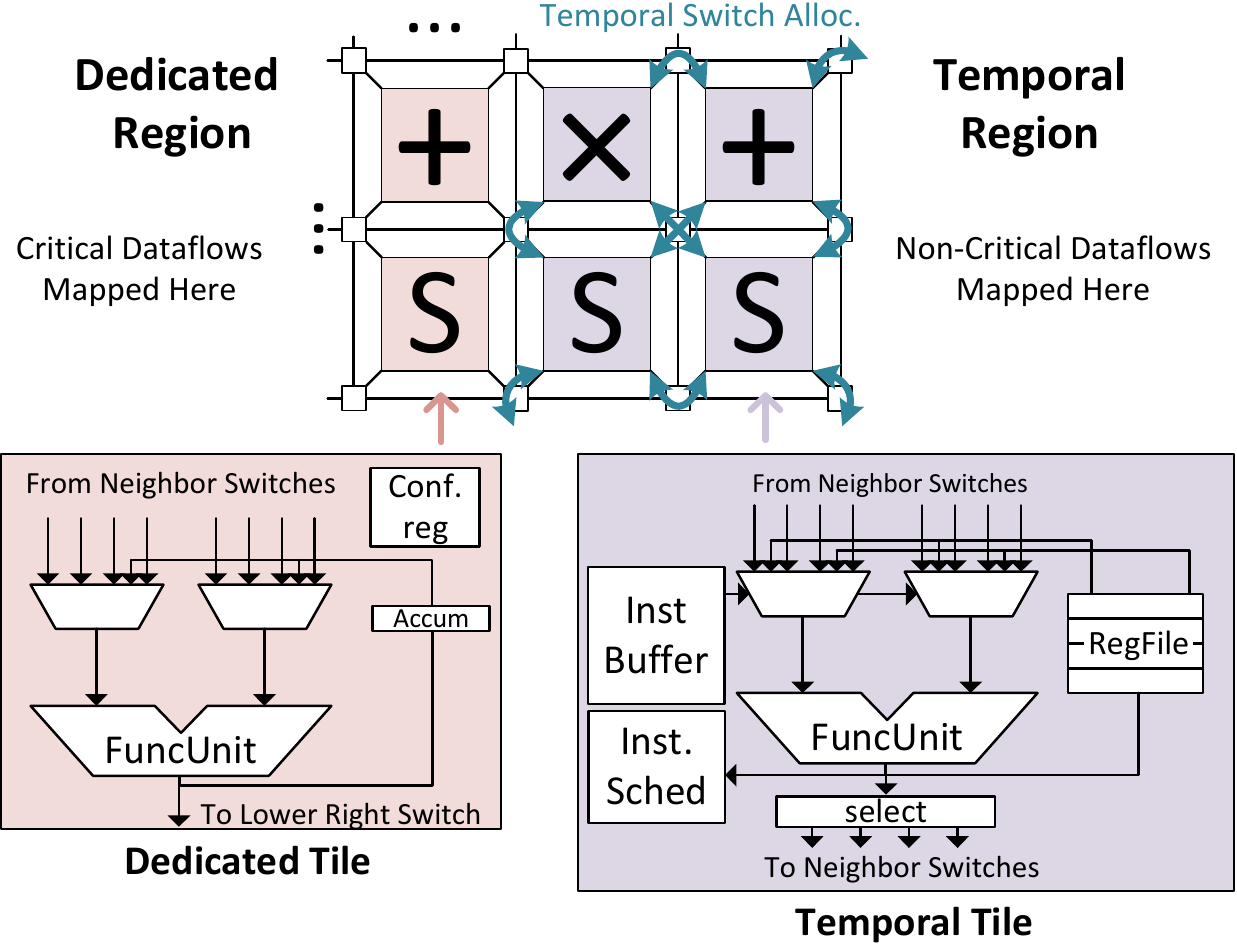}
  \vspace{-0.07in}  
  \captionof{figure}{Heterogeneous Computation Fabric and Tiles}
\label{fig:comp-fabric}
  \vspace{-0.05in}  
\end{figure}

There are two key types of fabrics with different tradeoffs. \textbf{Dedicated
fabrics} are those that restrict each execution resource (tile) to execute only
a single instruction, but pipelined at full throughput (eg. FPCA~{\cite{fpca}},
Q100~{\cite{q100}}, Tartan~{\cite{tartan}}, PipeRench~{\cite{piperench}}, and
DySER~{\cite{dyser}}).  In contrast, \textbf{temporal fabrics} are those that
may time-multiplex multiple different static instructions over the same
resources (eg.  TRIPS~{\cite{trips}}, WaveScalar~\cite{wavescalar}, 
Triggered Insts~{\cite{triggered-insts} and RAW~\cite{raw}}).\footnote{A VonNeumann core is also
\emph{temporal} in this context, but does not yield enough performance for this use in
REVEL.}  While dedicated fabrics only need to wait for the arrival of data
for dataflow execution, temporal fabrics require token matching, meaning more
complex structures (and power/area overhead).

Because critical dataflows are often easily vectorizable by the compiler, they can be scaled
to the size of the fabric, and be executed more efficiently on the more power/area efficient 
dedicated fabric.  However, non-critical 
dataflows often have \emph{many} instructions, and running them on a dedicated fabric would be
a waste of resources (eg. FP units) as they would attain poor utilization. 
Therefore, non-critical dataflows would be best run on a temporal fabric.  It would also be
inefficient to run critical dataflows on a (smaller) temporal fabric, as contention for resources would
degrade the throughput.  An over-provisioned temporal fabric can
alleviate this, at the cost of significant power/area overhead.

\paragraph{Criticality-Specialization}  Given the above, our approach is to make the
fabric heterogeneous: provision most of the fabric's resources to be a
dedicated fabric to enable fast execution of the critical dataflows, and
allocate a smaller portion to be a temporal fabric, which can execute
non-critical regions efficiently.  Figure~\ref{fig:comp-fabric} shows the lower
corner of REVEL's heterogeneous compute fabric, which embeds the
temporal fabric's network and compute into the dedicated fabric. 

The physical network for both fabrics is a circuit-switched mesh
with no flow control.  The dedicated tiles simply select inputs,
performs computations, and forward outputs in fully pipelined fashion
according to the dataflow configured by the mesh.
The dataflow compiler must equalize delays for each operand to ensure correct execution.


The temporal fabric is embedded within the circuit-switched
mesh, using a pattern shown by 
blue arrows in Figure~\ref{fig:comp-fabric}.  This allows temporal units
to communicate without interfering
with the dedicated region
(ie. no horizontal/vertical links consumed). 
The temporal tile microarchitecture is based on Triggered
Instructions~\cite{triggered-insts}, which performs operations based
on the arrival of data to a queue at the input or output.  A register
file holds live-state of waiting instructions.  


Note that dataflows communicate through ports (exiting and re-entering the
fabric).  The benefit of integration into the same 
network is that when there are no non-critical dataflows, the temporal
region may be reclaimed for use by critical instructions.
Section~\ref{sec:compiler} details compilation concerns.

\section{Generality of our approach} \label{sec:general}
Finally, we argue our approach is applicable
to other architectures.
Table~\ref{tab:other-arch} explains how to add FGOP capabilities to out-of-order (OOO)
cores and Plasticine~\cite{plasticine}, a reconfigurable
dataflow fabric, programmed using parallel patterns.  This table describes changes necessary
in the software, ISA and hardware.

\section{Software Stack} \label{sec:compiler}

A program is decomposed into two components: 1. C+intrinsics specifying the Von
Neumann control program, and 2. Dataflow specification which is mapped onto the
compute fabric.  A dataflow compiler (described in the next paragraph) is
responsible for producing the hardware configuration bits for the temporal and
dedicated portion of the fabric.  These are finally compiled together to create
the final RISCV binary.

\paragraph{Dataflow Compiler} We implemented a spatial architecture compiler 
(eg. \cite{wavescalar-sched,space-time-scheduler,edge-centric-scheduler,dresc,ilp-scheduler})
which maps computation and communication of all dataflows together on the compute fabric.
For the dedicated dataflows, all operand timing paths
must be equalized, and there is no sharing of resources.  For the temporal
dataflows, the goal is to minimize resource contention. Usually instructions
from a temporal/dedicated dataflow map to the corresponding region of the
compute fabric.  However, temporal instructions may map to the dedicated fabric
to minimize utilization, and dedicated instructions may be mapped to the
temporal fabric to minimize latency or network congestion, provided that there
are enough resources either way.  To balance these objectives, we take a
simulated annealing approach similar to the Pathfinder
algorithm~\cite{pathfinder} and prior stochastic
schedulers~\cite{hybrid-sched}, which allows resource over-provisioning to
determine and then constrain heavily needed network and execution resources.


\section{Evaluation Methodology} \label{sec:methodology}

\begin{table}[]
\footnotesize
\renewcommand{\arraystretch}{0.84}
\begin{tabular}{@{}p{0.25cm}p{0.7cm}p{1.85cm}p{4.55cm}@{}}
\toprule
\multirow{15}{*}{\rotatebox[origin=c]{90}{Revel Lane ($\times$ 8)}} & \multirow{4}{*}{CGRA}                         & PEs                                         & 14 add, and 3 sqrt/div, 9 mult   \\ 
                                                         &                                               &  Div/Sqrter                                            & Lat.: 12 Cyc., Thr.: 5 Cyc.\\
                                                         &                                               & SubwrdSIMD                                 & 4-way Fixed-point, 2-way FP                                             \\  
                                                         &                                               & \textbf{Data Firing}                         & \textbf{4 Independent Dataflows}                                  \\  
                                                         &                                               & \textbf{Temporal PE}                      & \textbf{2x1 (32 Insts/FU) }                           \\ \cmidrule(l){2-4} 
                                                         & \multirow{2}{0.75cm}{Vector Ports} & Width                                        & 2$\times$512, 2$\times$256, 1$\times$ 128, 1$\times$ 64 bit \\ 
                                                         &                                               & Capability                                   & 4-entry FIFO\textbf{+Config. Reuse}                           \\ \cmidrule(l){2-4} 
                                                         & \multirow{2}{*}{\parbox[t]{0.4cm}{Stream Control}}                      & SPAD Ctrl.                          & \textbf{Induct. Addr. Gen}, 8-Ent. Table   \\ 
                                                         &                                               & XFER Ctrl.                         & 8-Entry Stream-table                                             \\ 
                                                         &                                               & Cmd Queue                                    & 8-Entry Cmd Queue                                                       \\ \cmidrule(l){2-4} 
                                                         & \multirow{2}{*}{SPAD}                         & Structure                                    & 8Kb, Single-bank                                                        \\ 
                                                         &                                               & Bandwidth                                    & 512 Bits (1R/1W Port)                                                   \\   \cmidrule(l){2-4}
                                                         & \multirow{3}{*}{Net.}                     & SPAD-Ports           & 512 Bit Dedicated Bus                                                   \\  
                                                         &                                               & \textbf{XFER-Ports} & \textbf{512 Bit Dedicated Bus}                                          \\ 
                                                         &                                               & Ports-CGRA           & Point-to-Point 64-bit links                                             \\ \midrule
\rotatebox[origin=c]{90}{\parbox[t]{0.42cm}{Ctrl Core}~~}                                            & \multicolumn{3}{p{7cm}}{RISCV ISA~\cite{riscv}, 5-stage, single-issue, 16kb d\$, insts. added for stream-commands}                                                   \\ \midrule
\multirow{2}{*}{\rotatebox[origin=c]{90}{\parbox[t]{0.48cm}{Shr. SPD}~} }                                             &                                       \multicolumn{3}{p{7cm}}{Structure: 128Kb, Single-bank}                                                                                 \\  
                                                                                                              &  \multicolumn{3}{p{7cm}}{ Bandwidth: 512 Bits (1R/1W Port)}                                                                              \\ \midrule
\multirow{2}{*}{\rotatebox[origin=c]{90}{Net.}}                                & \multicolumn{3}{p{7cm}}{\textbf{Inter-lane: 512 Bit Data Bus (8-bit Cmd Sync)}}                                                         \\  
                                                                                                           & \multicolumn{3}{p{7cm}}{Shared scratchpad Bus: 512 Bit Shared Bus}                                                                          \\ \bottomrule
\end{tabular}
\vspace{-0.05in}
  \caption{REVEL Params (bold features for FGOP)} \label{tab:arch-param}
\vspace{-0.02in}
\end{table}


\paragraph{REVEL Modeling} REVEL hardware parameters are in Table~\ref{tab:arch-param}.
For performance, all blocks are modeled at a cycle level within a modified gem5~\cite{gem5}\cite{risc5}.
We synthesized a single lane of REVEL (heterogeneous 
fabric, stream control, ports, command queue, XFER unit)
using Synopsys DC, 28nm tech library. The design meets
timing at 1.25GHz.  An open source triggered instructions implementation 
was our reference for the temporal fabric~\cite{trig-impl}.
Results from synthesis, with Cacti 7.0~\cite{cacti} for SRAMs,
are used to create an event-based power model and area model.

\paragraph{ASIC Analytical Models}
These highly-optimistic models (Table~\ref{tab:asic-model})
are based on the optimized algorithms, and are 
only limited by the algorithmic critical path and
throughput constraints, with equivalent FUs to REVEL.
ASIC area and power models only count FU and scratchpad power.

\begin{table}[h]
\vspace{-0.1in}
\begin{center}
{
\setlength{\tabcolsep}{4pt}
\footnotesize
\begin{tabular}{llll}
\toprule
\multicolumn{2}{l}{SVD}                                            & QR                                     & MM                          \\
	$48m+2\text{QR}(n)+\lceil\frac{n^3}{4}\rceil$ & \multicolumn{2}{l}{$40n+n^2+\sum\limits_{i=1}^{n}(i+i*n)$} &  $\lceil\frac{nmp}{8}\rceil$\\
\midrule
	Solver                            & FFT                   & Cholesky                                                        & Centro-FIR                      \\
	2$\sum\limits_0^{n-1} \max(\lceil\frac{i}{4}\rceil, 14)$ & $\frac{n}{8}\log{n}$  & $\sum\limits_{i=1}^{n-1} \max (\lceil\frac{i^2}{4}\rceil, 24)$  & $\lceil\frac{n-m+1}{4}\rceil$  \\
\bottomrule
\end{tabular}
}
\end{center}
\vspace{-0.12in}
  \caption{Ideal ASIC Models \textnormal{(assumes FU lat. from Table~\ref{tab:arch-param}})}
\vspace{-0.12in}
\label{tab:asic-model}
\end{table}


\paragraph{Comparison Methodology}  
For fairness 
we compare designs with similar 
max. per-cycle throughput:
\begin{itemize} 
  \item \textbf{TI 6678 DSP (@1.25GHz)} 8-core DSP,
    each core has 16-FP adders/multipliers, using DSPLIB\_C66x\_3.4.0.0.  
 \item \textbf{OOO Core: Intel Xeon 4116 (@2.1GHz)}  Conventional OOO processor
   using highly-optimized Intel MKL library.  (8 cores used)
  \item \textbf{REVEL-No-FGOP:}  REVEL without 
    FGOP support (8 Lanes). To evaluate, we therefore also implement highly-optimized non-FGOP workload versions.
\end{itemize}

\begin{table}[tb]
{ \footnotesize
\begin{center}
\setlength{\tabcolsep}{1.8pt}
\begin{tabular}{llllllll}
\toprule
Workload & Data Size                            & Lane    & Acc     & Dep  & Reuse   & Het  & Vec \\
\midrule                                                                             
SVD       & \textbf{12},16,24,\textbf{32}       & 1       & RI       & Y   & Y      & Y    & Y \\
QR        & \textbf{12},16,24,\textbf{32}       & 8       & RI       & Y   & Y      & Y    & Y \\
Cholesky  & \textbf{12},16,24,\textbf{32}       & 8       & RI       & Y   & Y      & Y    & Y \\
Solver    & \textbf{12},16,24,\textbf{32}       & 1       & RI       & Y   & Y      & Y    & Y \\
FFT       & \textbf{64},128...\textbf{1024}     & 1       & RR       & N   & Y      & N    & N \\
GEMM      & \textbf{12},24,\textbf{48}x16x64    & 8       & RR       & N   & Y        & N    & N \\
FIR       & \textbf{12},16,24,\textbf{32}       & 8       &  I       & N   & Y        & N    & N \\
\bottomrule
\end{tabular}
\end{center}
}
\vspace{-0.19in}
  \caption{Workload Params. and FGOP Features; small and large 
  \textnormal{sizes bolded;
  Lane: \#Lanes in latency ver.,  Acc: Access pat.
  Dep: Fine-grain deps, Reuse: stream-reuse, Het: Heterog. fabric, Vec: implicit vect. masking}}
\label{tab:params}
\vspace{-0.05in}
\end{table}

\paragraph{Workload Versions}
We make comparisons in two different settings, high-throughput and low-latency.
The throughput setting assumes there exist multiple data items to parallelize over,
while the latency setting assumes only one.
We implement both throughput and latency optimized REVEL workloads,
where latency-optimized spreads work across multiple lanes.
Throughput versions use each lane in data-parallel fashion. Note that we could not
profitably parallelize any FGOP kernel across multiple DSP/OOO
cores, even using native libraries, so their latency-optimized versions only
use a single core. Table~\ref{tab:params}
describes data-sizes, and also how FGOP features were used by each workload.


\section{Evaluation} \label{sec:eval}

We broadly answer the question of whether fine-grain ordered
parallelism is exploitable in DSP workloads, and if REVEL's
execution model, architecture, and microarchitecture is effective. 
What we find overall is that REVEL's ability to exploit FGOP 
leads to order-of-magnitude speedup and area-normalized
performance over traditional DSPs.

We first discuss the applicability
of FGOP features and overall latency and throughput potential.
We then explain how performance improvements were achieved by analyzing
cycle-level bottleneck breakdowns, and incremental performance improvements.
We also answer the question of sensitivity to temporal region size and 
address-generation capability. 
Finally, we analyze the area and power breakdowns, comparison of normalized
performance, and compare to optimistic ASIC models.

\paragraph{Q1. Can workloads use REVEL's FGOP features?}
Table~\ref{tab:params} shows the applicability of FGOP features. Matrix factorization/decomposition 
workloads (QR, SVD, Cholesky, Solver) use all FGOP features.  Even
non-FGOP workloads took advantage of streaming-reuse to reduce scratchpad bandwidth,
and FIR had a short inductive access phase.

\begin{figure}[t]
\includegraphics[width=\linewidth]{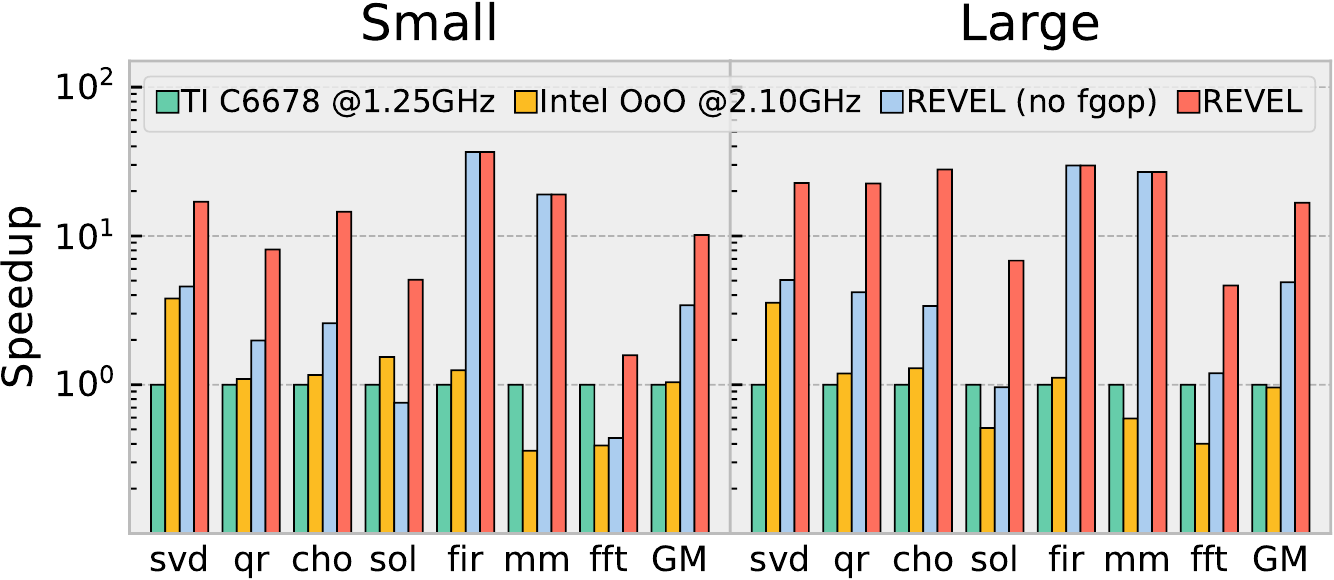}
\vspace{-0.18in}
\caption{Latency-optimized kernel performance}
\vspace{-0.12in}
\label{fig:perf}
\end{figure}

\begin{figure}[t]
\includegraphics[width=\linewidth]{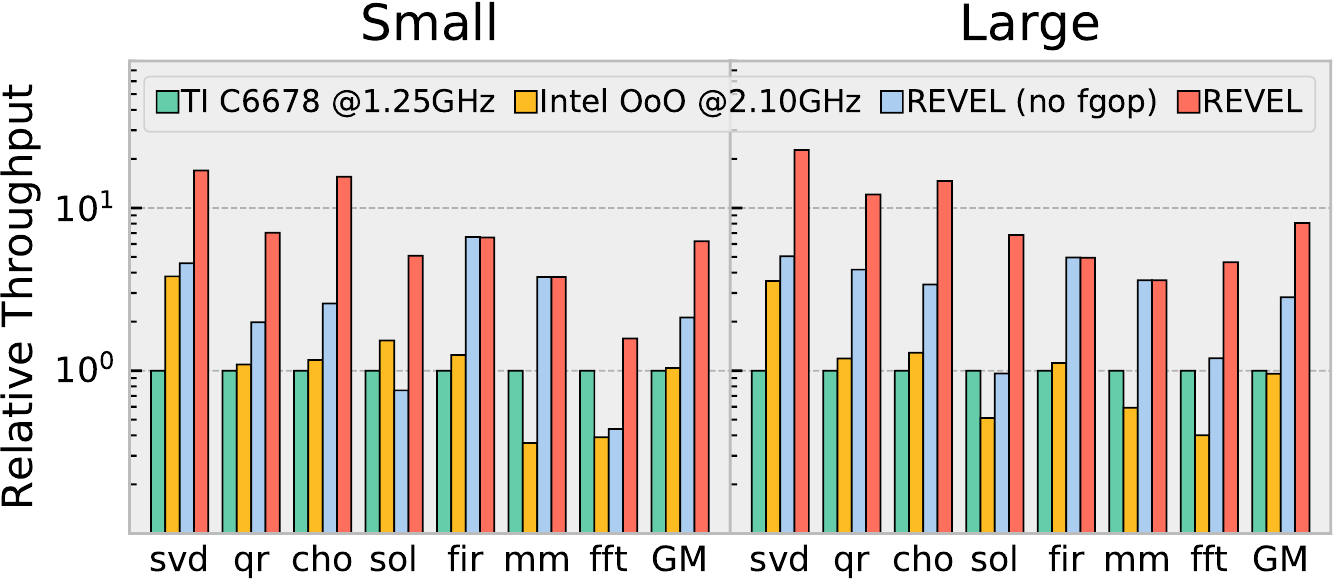}
\vspace{-0.21in}
\caption{Throughput-optimized kernel performance}
\vspace{-0.12in}
\label{fig:tput}
\end{figure}

\paragraph{Q2. How much speedup do REVEL's execution model and FGOP features provide?} 
The speedups over DSP for latency optimized
codes are shown in Figure~\ref{fig:perf} for both
small and large matrices. The DSP and CPU have similar mean performance.
REVEL attains up to 37$\times$ speedup, with geomean of 10$\times$ and 17$\times$ 
for small and large data sizes.
Considering just workloads which exhibit FGOP, 
the speedup from FGOP specialization is 6.1$\times$ (large size). 
The benefit of REVEL's dataflow/vector-stream model without FGOP provides 2.8$\times$ speedup over DSP.
The DSP is only competitive on the small FFT, as REVEL here requires
 multiple-configurations.

Performance for throughput-optimized kernels (data parallelism across lanes), 
is shown in Figure~\ref{fig:tput}.  For small and large sizes, REVEL gets a speedup of
6.3$\times$ and 8.1$\times$ over the DSP and CPU.
Again, considering just workloads which exhibit FGOP, 
the speedup from FGOP specialization is 4.4$\times$ (large size).
REVEL's dataflow/vector-stream model provides the other 2.6$\times$ speedup over the DSP.
The performance tradeoffs here are similar, except the advantage of parallelizing
across lanes is diminished due to data-parallel execution.  

\emph{The vector-stream control and FGOP-exploitation enables combined
order-of-magnitude speedups.}

\begin{figure}[t]
\includegraphics[width=\linewidth]{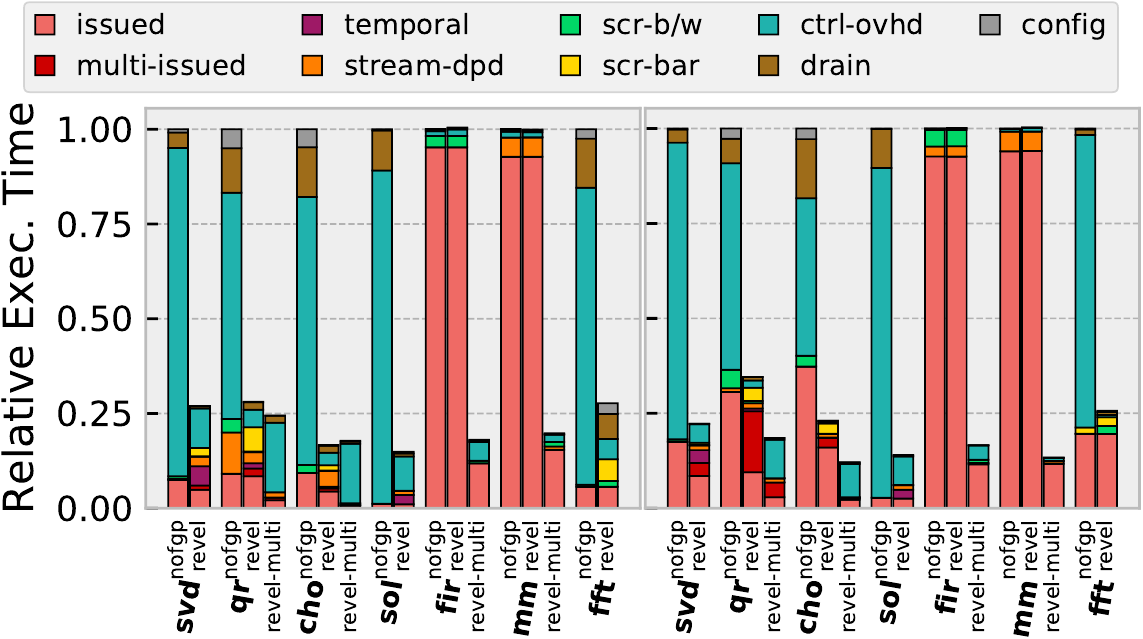}
\vspace{-0.21in}
\caption{REVEL's Cycle-level bottlenecks} \label{fig:bottle}
\vspace{-0.1in}
\end{figure}

\paragraph{Q3. Why does exploiting FGOP help REVEL?}  
Figure~\ref{fig:bottle} overviews REVEL's cycle-level behavior, 
normalized to non-FGOP hardware.
Latency-optimized workloads are labeled as ``multi''.
To explain the categories, \emph{issue} and \emph{multi-issue} means that
one or multiple dedicated dataflow fired, and \emph{temporal} means only a temporal dataflow
fired during that cycle.  All other categories represent overhead, including the \emph{drain}
of the dedicated fabric, \emph{scr-b/w} and \emph{scr-barrier} for bandwidth
and synchronization, \emph{stream-dpd} for waiting on dependences, and
\emph{ctrl-ovhd} for waiting on the control core. 

The clearest trend is that exploiting FGOP reduces the
control overhead for both small and large matrix sizes.  Also,
exploiting FGOP enables parallelism between dataflows, which can be
seen in the multi-issued category, especially prevalent for the larger matrix
sizes of FGOP kernels.

\emph{Exploiting FGOP increases parallelism and reduces control overhead,
enabling higher hardware utilization.}

\begin{figure}[t]
\includegraphics[width=\linewidth]{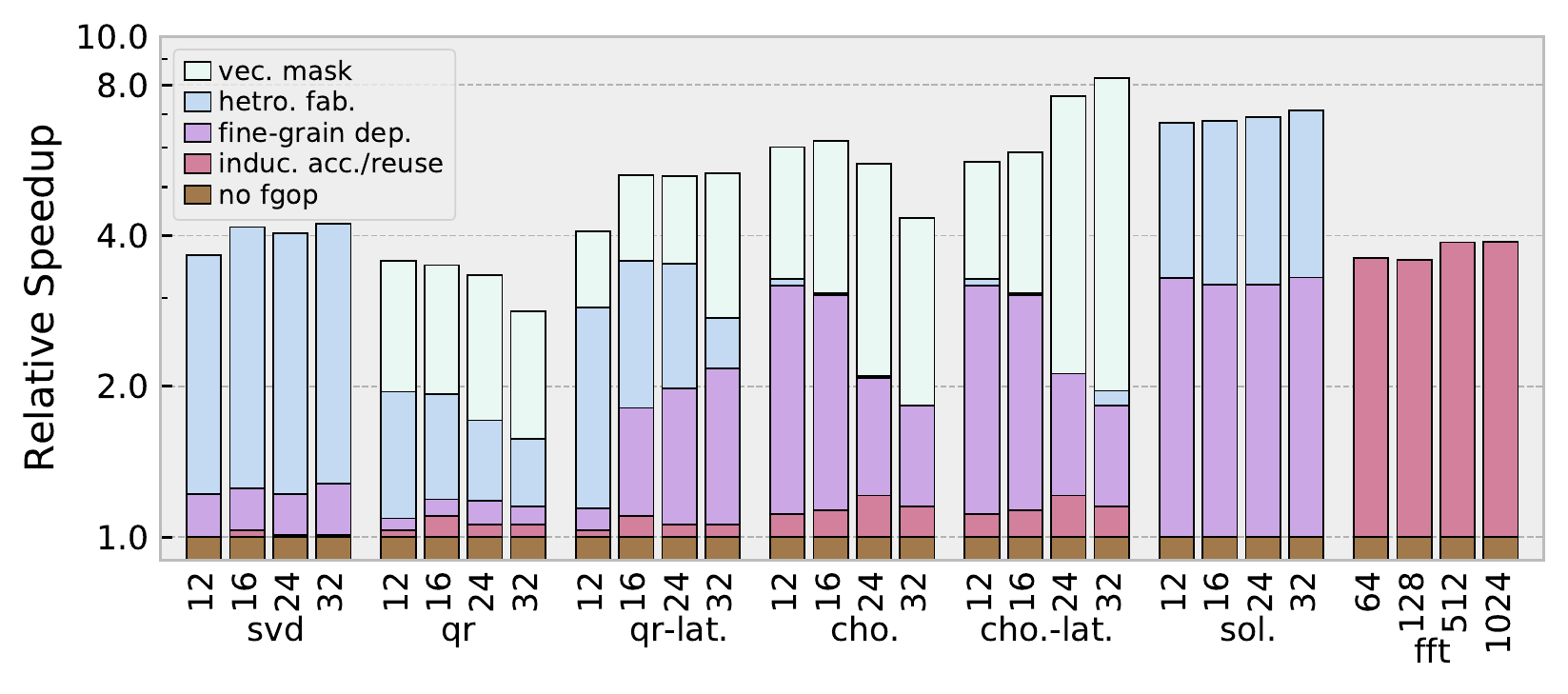}
\vspace{-0.27in}
\caption{Performance Impact of Each Mechanism.}
\vspace{-0.08in}
\label{fig:inc}
\end{figure}

\paragraph{Q4. What is the impact of each mechanism?}
Figure~{\ref{fig:inc}} shows the incremental speedup of each hardware/software
feature (so 5 versions of each kernel).  Latency-optimized versions have ``lat'' as a suffix.

The inductive benefit is small alone, as it reduces control but does not increase parallelism.
Only FFT benefits greatly by using inductive reuse to reduce SPAD bandwidth. 
Most workloads were accelerated dramatically from exploiting fine-grain dependences.
However, QR and SVD have complex sub-critical regions,
so they only see the benefit after adding the heterogeneous fabric.
Throughput-optimized QR suffers from local memory access because of the shrinking matrix sizes,
but latency-optimized QR converts these to inter-lane data streams.
Solver is also accelerated by the heterogeneous fabric because it is latency sensitive,
and collapsing sub-critical instructions can reduce latency.
Cholesky's triangular access implies large gains from implicit
vector masking.

\emph{REVEL's mechanisms together enable high performance.}


\paragraph{Q5. What is the biggest remaining overhead for REVEL?}
As shown in Figure~\ref{fig:bottle}, this is the drain time on smaller workloads,
often caused by reconfiguration.  
This is more of an issue for the smaller matrices and especially FFT, 
where the datapath should be reconfigured for each algorithm phase. 
\emph{REVEL's reliance on deep pipelines causes 
config/serialization penalty on extremely short phases.}

\paragraph{Q6. What are the sources of area and power?}
Table~\ref{tab:powerarea} shows the breakdown; the largest
source (especially power) comes mostly from the floating point
units.  At 28nm, REVEL is 1.79mm$^2$.  Note that the
control core is now only about one 50th of the overall area.

\begin{table}[t]
\footnotesize
\begin{center}
\begin{tabular}{p{1.3cm}lrr}
\toprule
                               &                                         & area(mm$^2$)  & power(mw) \\
\midrule
       \multirow{3}{*}{\makecell{Compute\\Fabric}}   & Dedi. Net. (23)  & 0.05          & 71.40      \\
~                                                    & Temp. Net. (2)   & 0.01          & 14.81      \\
~                                                    & Func. Units      & 0.07          & 74.04      \\

                               \cmidrule{2-4}
~                                                    & Total Fabric     & 0.13          & 160.25     \\
\midrule
\multicolumn{2}{l}{Control \small{(ports/XFER/str. ctrl)}}              & 0.03          & 62.92     \\
\multicolumn{2}{l}{SPAD-8KB}                                            & 0.06          & 4.64      \\
\midrule
\multicolumn{2}{l}{\textbf{1 Vector Lane}}                              & 0.22          & 207.90    \\
\multicolumn{2}{l}{Control Core}                                        & 0.04          & 19.91     \\
\multicolumn{2}{l}{REVEL}                                               & 1.79          & 1663.3    \\
\bottomrule

\end{tabular}
  \end{center}
  \vspace{-0.17in}
  \caption{Area and Power Breakdown (28nm)}
  \vspace{-0.13in}
\label{tab:powerarea}
\end{table}


\paragraph{Q7. Does REVEL have better perf/mm$^2$?}
REVEL's high speedup with only small area overhead (over the DSP) for
the computation fabric's networks results in a large performance/mm$^2$ advantage:
1308$\times$ over the OOO core and 8.3$\times$ over the DSP.

\begin{figure}[t]
\begin{center}
\includegraphics[width=0.694\linewidth]{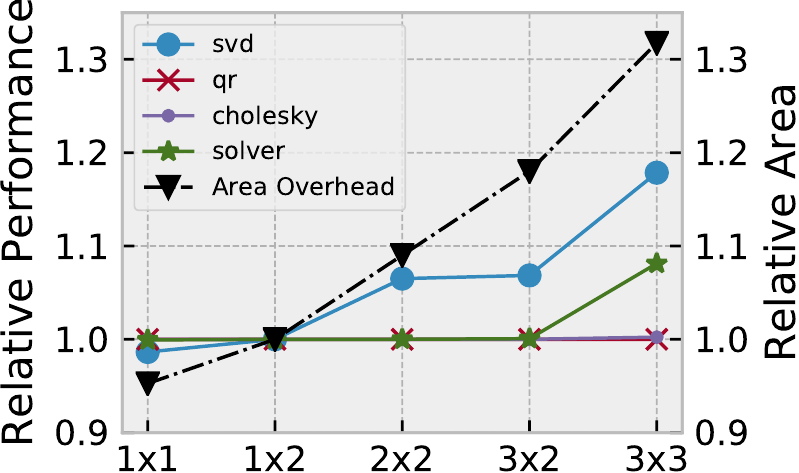}
  \caption{Temporal region sensitivity (width$\times$height) }
\vspace{-0.14in}
\label{fig:temporal-sens}
\vspace{-0.08in}
\end{center}
\end{figure}

\paragraph{Q8. How sensitive is REVEL's performance to the size of the temporal region?}
Because temporal tiles cost more than 5$\times$ the area than dedicated tiles (dedicated tile: 2265$\mu$m$^2$, 
temporal tile: 12062$\mu$m$^2$), it is important to choose the correct temporal region size.
Figure~\ref{fig:temporal-sens} shows REVEL's performance sensitivity to this size, as well as the
area tradeoff.
SVD and QR have the largest regions, so are affected the most, but a 1$\times$1
temporal region only has 13\% overhead. We choose this size to minimize area penalty.



\paragraph{Q9. Do we require a heterogeneous fabric?} \label{q9}
To create a purely dedicated fabric, we would have had to support 52 additional
dedicated tiles for our largest temporal region in SVD, costing about 2.75$\times$ fabric area.
Similarly, for the entire design to be temporal, it would have cost around 2.5$\times$ fabric
area.
\emph{A heterogeneous fabric provides the best performance/area ratio.}

\paragraph{Q10. Would even more complex inductive streams have reduced control overhead?} \label{q10}
Our analysis so far has shown that using 2D inductive streams can reduce control overhead
and improve performance significantly.  An interesting question is whether supporting
higher dimension stream-access could have helped further.

\begin{figure} [t]
\begin{center}
\includegraphics[width=\linewidth]{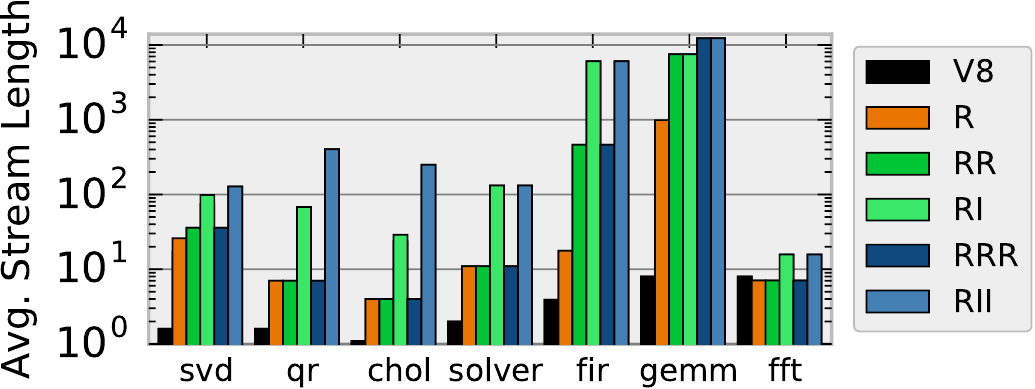}
\end{center}
  \vspace{-0.21in}
  \caption{Stream-type Access Length Comparison}
\label{fig:acc-len}
\end{figure}

\begin{figure} [t]
\begin{center}
\includegraphics[width=\linewidth]{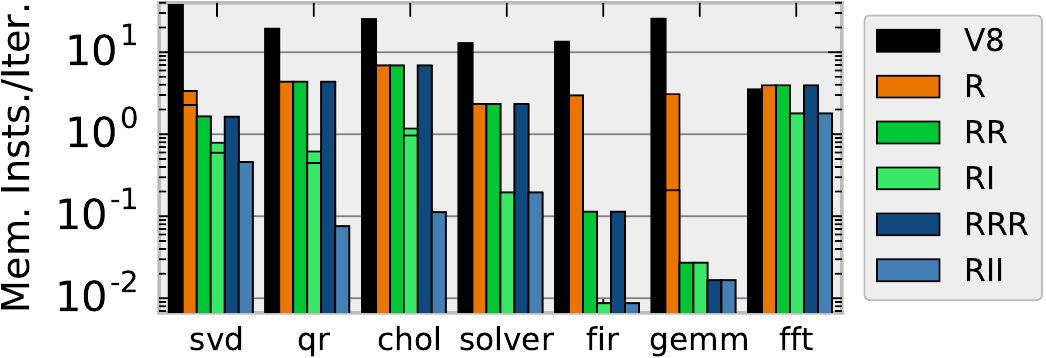}
\end{center}
  \vspace{-0.21in}
  \caption{Control overhead of various Capabilities measured in control instructions
  per iteration.  The stacked bar indicates
  additional control overhead if stream-reuse technique is disabled.}
\label{fig:acc-iters}
  \vspace{-0.08in}
\end{figure}

To analyze stream capabilities analytically, we implement a static compiler analysis
in LLVM~\cite{llvm}, using scalar evolution analysis~\cite{scev} for the closed-form representation 
of address patterns and loop termination with respect to induction variables. This analysis
can determine the length of a given stream for each pattern.
Figure~\ref{fig:acc-len} shows the average length of a stream: 
the number of loop iterations the pattern describes.  
We also calculate the number of effective memory instructions
per inner-loop iteration, ``Mem. Insts/Iter'', which is a measure of the control
overhead (Figure~\ref{fig:acc-iters}). We consider
vector (V), 1D streams (R), 2D streams (RR and inductive RI) and 3D streams (RRR and inductive RII).

Regular workloads like GEMM require only a
low dimension rectangular access pattern for a long length.  However,
FGOP workloads show much higher lengths \emph{only} with inductive 
access capability (RI or RII capability).  
This benefit translates to fewer memory instructions
per-iteration.  A value of less than 1 in Figure~\ref{fig:acc-iters} means that
fewer than one control instructions would need to be issued per cycle.  This
helps to explain \emph{why} vector instructions alone are insufficient for
parallelism -- because they require too much specification of work.
Fortunately, the RI capability always either achieves a control
overhead below 1 inst/iter or matches the least overhead capability.

The ability to reuse stream values as part of the stream definition can also
reduce control overhead.  The control overhead if this feature is disabled
is shown by the stacked bar in Figure~\ref{fig:acc-iters}.  This
benefit is modest; the more important reason for stream-reuse is to
reduce memory bandwidth. 

\emph{2D Induction streams (RI) are necessary to reduce control overhead, 
and RII may provide only a small energy advantage, but is also more complex.}

\paragraph{Q11. How competitive is REVEL with custom ASICs?}
Table~\ref{tab:comp-asic} shows the performance-normalized power and area overhead, as compared
to ASIC analytical models. 
REVEL is mean 2.2$\times$ power.  This is mostly due to the
control logic (vector ports, bus, and etc.) and reconfigurable networks.
It is 0.55$\times$ the area of the combined ASIC.
Note this is highly optimistic for ASICs, as the performance model assumes perfect
pipelining, and the power model assumes no control.

\setcounter{table}{5}
\begin{table}[H]
\vspace{-0.05in}
\begin{center}
{
\large
\setlength{\tabcolsep}{3pt}
\renewcommand{\arraystretch}{0.84}
\footnotesize
\begin{tabular}{lllllllll}
\toprule

Workloads   & SVD & QR  & Cho.& Sol.& FIR & MM   & FFT & Mean \\
\midrule
Power Ovhd. & 3.5 & 2.1 & 2.2 & 2.0 & 2.0 & 1.9  & 1.9 & 2.2  \\
\midrule
Area  Ovhd. & 3.8 & 2.4 & 2.5 & 2.7 & 2.2 & 2.1  & 2.6 & 2.6/0.55 \\
\bottomrule
\end{tabular}
}
\end{center}
  \vspace{-0.16in}
  \caption{Power/Area overheads to ideal ASIC (iso-perf)}
\label{tab:comp-asic}
  \vspace{-0.12in}
\end{table}

\emph{REVEL is competitive with ASICs, and could replace 
fixed-function accelerators or conventional DSPs in some designs.}

\vspace{-0.1in}

\section{Related work} \label{sec:related}


\paragraph{DSP Accelerators}  
Many application/domain-specific reconfigurable designs have
targeted DSP algorithms.  Fasthuber et.
al~\cite{energy-eff-comm} outline the basic approaches.  One representative
example includes LAC~\cite{mat-fact}, targeted at matrix factorization.

\paragraph{Ordered Parallelism and Synchronization}
A conceptually similar work to ours from the GPGPU space is
dMT-CGRA~\cite{dmt_cgra}, which proposes inter-thread communication
between SIMT threads~\cite{sgmf,mt-cgrf}.
Prabhakar et al~\cite{pp-fpga}
develops ``nested-parallelism,'' which enables coupling of
datapaths with different nesting of parallel patterns.
Swarm~\cite{swarm} also targets a form of ordered parallelism by building abstractions
on top of a task-parallel model, targeting irregular data-dependent parallelism~\cite{tao}.

\paragraph{Task-parallel+Acceleration}  An alternative model to
ours is task-based parallelism plus some form of acceleration 
to reduce the synchronization overhead (eg. TAPAS~\cite{tapas}).
Task parallelism has the benefit of dynamic load balancing, but this
does not appear to be necessary our DSP workloads.

\paragraph{Flexible Vectorization}
Our vector-stream control paradigm is inspired by
prior techniques which marshal independent execution lanes to create a vector-like
execution when useful.  This includes Vector-Threading~\cite{mavenvt,scalevt}, 
Vector-Lane Threading~\cite{vlt}, and Loop-Task Accelerators~\cite{loop-task}.
REVEL also marshals lanes to reduce control and increase parallelism, 
but its lanes are autonomous once programmed with streams.

Some techniques apply vectorization with 
reconfigurability, eg. Libra~\cite{libra} 
and DySER~\cite{dyser}, which can create/reconfigure
vector lanes. REVEL also amortizes control through time. 

\begin{wrapfigure}{r}{1.7in}
  \vspace{-0.15in}
  \renewcommand{\arraystretch}{0.84}
  \small
  \setlength{\tabcolsep}{0.04in}
\begin{tabular}{@{}lr@{}}
  \toprule
  \multicolumn{2}{c}{\textbf{Address Gen Capability}} \\
  Name                                           & Type \\ \midrule
  Imagine~\cite{imagine}                         & R \\
  Q100~\cite{q100}                               & R \\
  Accel DMA~\cite{shao}                          & R \\
  Softbrain~\cite{stream-dataflow}               & RR \\
  RSVP~\cite{rsvp}                               & RR \\
  CoRAM++~\cite{corampp}                         & RR \\ 
  APMC~\cite{apmc}                               & RR \\
  \textbf{REVEL}                                 & \textbf{RI~} \\
  FPCA~\cite{fpca}                               & RRR \\ 
  \bottomrule
\end{tabular}
\end{wrapfigure}

\paragraph{Stream-based ISAs}
Several prior architectures have stream primitives.
We list their address capability as compared to REVEL on the right.  
We believe REVEL is the only one to support inductive patterns.




\section{Discussion and Conclusion} \label{sec:conc}

This paper identified fine-grain ordered parallelism as
a common property across a variety of linear-algebra and DSP algorithms.
It is extremely difficult
to take advantage of using existing VonNeumann Vector and multi-threading architectures.
This work identified a set of abstractions and developed an execution model and hardware
implementation (REVEL) which could exploit this form of parallelism.


Our REVEL implementation was more than an order of magnitude lower latency (10$\times$-17$\times$),
and its performance per mm$^2$ was 6.7$\times$ that of a DSP (up to 16$\times$).  
Overall, REVEL's design offers large advantages over existing architectures for
important signal processing workloads, and is a promising alternative to existing DSPs and beyond.  


\balance


\bibliographystyle{ieeetr}
\bibliography{ref}

\begin{thebibliography}{10}

\bibitem{mobile-edge-whitepaper}
Y.~C. Hu, M.~Patel, D.~Sabella, N.~Sprecher, and V.~Young, ``Mobile edge
  computing---a key technology towards 5g,'' {\em ETSI white paper}, vol.~11,
  no.~11, pp.~1--16, 2015.

\bibitem{stream-dataflow}
T.~Nowatzki, V.~Gangadhar, N.~Ardalani, and K.~Sankaralingam, ``Stream-dataflow
  acceleration,'' in {\em Proceedings of the 44th Annual International
  Symposium on Computer Architecture}, ISCA '17, (New York, NY, USA),
  pp.~416--429, ACM, 2017.

\bibitem{lssd}
T.~Nowatzki, V.~Gangadhar, K.~Sankaralingam, and G.~Wright, ``Pushing the
  limits of accelerator efficiency while retaining programmability,'' in {\em
  2016 IEEE International Symposium on High Performance Computer Architecture
  (HPCA)}, pp.~27--39, March 2016.

\bibitem{plasticine}
R.~Prabhakar, Y.~Zhang, D.~Koeplinger, M.~Feldman, T.~Zhao, S.~Hadjis,
  A.~Pedram, C.~Kozyrakis, and K.~Olukotun, ``Plasticine: A reconfigurable
  architecture for parallel paterns,'' in {\em Proceedings of the 44th Annual
  International Symposium on Computer Architecture}, ISCA '17, (New York, NY,
  USA), pp.~389--402, ACM, 2017.

\bibitem{morphosys}
H.~Singh, M.-H. Lee, G.~Lu, N.~Bagherzadeh, F.~J. Kurdahi, and E.~M.~C. Filho,
  ``Morphosys: An integrated reconfigurable system for data-parallel and
  computation-intensive applications,'' {\em IEEE Trans. Comput.}, vol.~49,
  pp.~465--481, May 2000.

\bibitem{dyser}
V.~Govindaraju, C.-H. Ho, T.~Nowatzki, J.~Chhugani, N.~Satish,
  K.~Sankaralingam, and C.~Kim, ``Dyser: Unifying functionality and parallelism
  specialization for energy-efficient computing,'' {\em IEEE Micro}, vol.~32,
  pp.~38--51, Sept. 2012.

\bibitem{beamforming}
R.~Mudumbai, G.~Barriac, and U.~Madhow, ``On the feasibility of distributed
  beamforming in wireless networks,'' {\em IEEE Transactions on Wireless
  communications}, vol.~6, no.~5, 2007.

\bibitem{polyphase}
H.~Johansson {\em et~al.}, ``Polyphase decomposition of digital
  fractional-delay filters,'' {\em IEEE signal processing letters}, vol.~22,
  no.~8, pp.~1021--1025, 2015.

\bibitem{centro-symm}
R.~Zhao, ``Wls design of centro-symmetric 2-d fir filters using matrix
  iterative algorithm,'' in {\em 2015 IEEE International Conference on Digital
  Signal Processing (DSP)}, pp.~34--38, July 2015.

\bibitem{half-band}
F.~Mintzer, ``On half-band, third-band, and nth-band fir filters and their
  design,'' {\em IEEE Transactions on Acoustics, Speech, and Signal
  Processing}, vol.~30, pp.~734--738, Oct 1982.

\bibitem{speech-svd}
M.~Dendrinos, S.~Bakamidis, and G.~Carayannis, ``Speech enhancement from noise:
  A regenerative approach,'' {\em Speech Communication}, vol.~10, no.~1,
  pp.~45--57, 1991.

\bibitem{qr-dsp}
D.~Patel, M.~Shabany, and P.~G. Gulak, ``A low-complexity high-speed qr
  decomposition implementation for mimo receivers,'' in {\em 2009 IEEE
  International Symposium on Circuits and Systems}, pp.~33--36, May 2009.

\bibitem{cholesky-dsp}
P.~Salmela, A.~Happonen, T.~Jarvinen, A.~Burian, and J.~Takala, ``Dsp
  implementation of cholesky decomposition,'' in {\em Joint IST Workshop on
  Mobile Future, 2006 and the Symposium on Trends in Communications. SympoTIC
  '06.}, pp.~6--9, June 2006.

\bibitem{lmmse-wcdma}
P.~Darwood, P.~Alexander, and I.~Oppermann, ``Lmmse chip equalisation for 3gpp
  wcdma downlink receivers with channel coding,'' in {\em ICC 2001. IEEE
  International Conference on Communications. Conference Record (Cat.
  No.01CH37240)}, vol.~5, pp.~1421--1425 vol.5, 2001.

\bibitem{wirelessdsp}
D.~Tse and P.~Viswanath in {\em Fundamentals of Wireless Communication}, New
  York, NY, USA: Cambridge University Press, 2005.

\bibitem{polybench}
L.-N. Pouchet, ``Polybench: The polyhedral benchmark suite,'' {\em URL:
  http://www. cs. ucla. edu/pouchet/software/polybench}, 2012.

\bibitem{llvm}
C.~Lattner and V.~Adve, ``{LLVM}: A compilation framework for lifelong program
  analysis \& transformation,'' in {\em CGO '04}, pp.~75--88.

\bibitem{ompchol}
A.~Buttari, ``Multicore and multicore programming with openmp,'' {\em URL:
  http://buttari. perso. enseeiht. fr/stuff/crgc\_mcore. pdf}, 2012.

\bibitem{imagine}
S.~Rixner, W.~J. Dally, U.~J. Kapasi, B.~Khailany, A.~L\'{o}pez-Lagunas, P.~R.
  Mattson, and J.~D. Owens, ``A bandwidth-efficient architecture for media
  processing,'' in {\em Proceedings of the 31st Annual ACM/IEEE International
  Symposium on Microarchitecture}, MICRO 31, (Los Alamitos, CA, USA),
  pp.~3--13, IEEE Computer Society Press, 1998.

\bibitem{rsvp}
S.~Ciricescu, R.~Essick, B.~Lucas, P.~May, K.~Moat, J.~Norris, M.~Schuette, and
  A.~Saidi, ``The reconfigurable streaming vector processor (rsvp),'' in {\em
  Proceedings of the 36th Annual IEEE/ACM International Symposium on
  Microarchitecture}, MICRO 36, (Washington, DC, USA), pp.~141--, IEEE Computer
  Society, 2003.

\bibitem{q100}
L.~Wu, A.~Lottarini, T.~K. Paine, M.~A. Kim, and K.~A. Ross, ``Q100: The
  architecture and design of a database processing unit,'' in {\em Proceedings
  of the 19th International Conference on Architectural Support for Programming
  Languages and Operating Systems}, ASPLOS '14, (New York, NY, USA),
  pp.~255--268, ACM, 2014.

\bibitem{corampp}
G.~Weisz and J.~C. Hoe, ``Coram++: Supporting data-structure-specific memory
  interfaces for fpga computing,'' in {\em 25th International Conference on
  Field Programmable Logic and Applications (FPL)}, pp.~1--8, Sept 2015.

\bibitem{apmc}
T.~Hussain, O.~Palomar, O.~Unsal, A.~Cristal, E.~Ayguadé, and M.~Valero,
  ``Advanced pattern based memory controller for fpga based hpc applications,''
  in {\em 2014 International Conference on High Performance Computing
  Simulation (HPCS)}, pp.~287--294, July 2014.

\bibitem{fpca}
J.~Cong, H.~Huang, C.~Ma, B.~Xiao, and P.~Zhou, ``A fully pipelined and
  dynamically composable architecture of cgra,'' in {\em Field-Programmable
  Custom Computing Machines (FCCM), 2014 IEEE 22nd Annual International
  Symposium on}, pp.~9--16, IEEE, 2014.

\bibitem{scalevt}
R.~Krashinsky, C.~Batten, M.~Hampton, S.~Gerding, B.~Pharris, J.~Casper, and
  K.~Asanovic, ``The vector-thread architecture,'' in {\em Proceedings of the
  31st Annual International Symposium on Computer Architecture}, ISCA '04,
  (Washington, DC, USA), pp.~52--, IEEE Computer Society, 2004.

\bibitem{mavenvt}
Y.~Lee, R.~Avizienis, A.~Bishara, R.~Xia, D.~Lockhart, C.~Batten, and
  K.~Asanovi\'{c}, ``Exploring the tradeoffs between programmability and
  efficiency in data-parallel accelerators,'' in {\em Proceedings of the 38th
  Annual International Symposium on Computer Architecture}, ISCA '11, (New
  York, NY, USA), pp.~129--140, ACM, 2011.

\bibitem{vlt}
S.~Rivoire, R.~Schultz, T.~Okuda, and C.~Kozyrakis, ``Vector lane threading,''
  in {\em 2006 International Conference on Parallel Processing (ICPP'06)},
  pp.~55--64, Aug 2006.

\bibitem{tartan}
M.~Mishra, T.~J. Callahan, T.~Chelcea, G.~Venkataramani, S.~C. Goldstein, and
  M.~Budiu, ``Tartan: Evaluating spatial computation for whole program
  execution,'' in {\em Proceedings of the 12th International Conference on
  Architectural Support for Programming Languages and Operating Systems},
  ASPLOS XII, (New York, NY, USA), pp.~163--174, ACM, 2006.

\bibitem{piperench}
S.~Goldstein, H.~Schmit, M.~Moe, M.~Budiu, S.~Cadambi, R.~Taylor, and
  R.~Laufer, ``Piperench: a coprocessor for streaming multimedia
  acceleration,'' in {\em Computer Architecture, 1999. Proceedings of the 26th
  International Symposium on}, 1999.

\bibitem{trips}
D.~Burger, S.~W. Keckler, K.~S. McKinley, M.~Dahlin, L.~K. John, C.~Lin, C.~R.
  Moore, J.~Burrill, R.~G. McDonald, W.~Yoder, and t.~T. Team, ``Scaling to the
  end of silicon with edge architectures,'' {\em Computer}, vol.~37,
  pp.~44--55, July 2004.

\bibitem{wavescalar}
S.~Swanson, K.~Michelson, A.~Schwerin, and M.~Oskin, ``Wavescalar,'' in {\em
  Proceedings of the 36th Annual IEEE/ACM International Symposium on
  Microarchitecture}, MICRO 36, (Washington, DC, USA), pp.~291--, IEEE Computer
  Society, 2003.

\bibitem{triggered-insts}
A.~Parashar, M.~Pellauer, M.~Adler, B.~Ahsan, N.~Crago, D.~Lustig, V.~Pavlov,
  A.~Zhai, M.~Gambhir, A.~Jaleel, R.~Allmon, R.~Rayess, S.~Maresh, and J.~Emer,
  ``Triggered instructions: A control paradigm for spatially-programmed
  architectures,'' in {\em Proceedings of the 40th Annual International
  Symposium on Computer Architecture}, ISCA '13, (New York, NY, USA),
  pp.~142--153, ACM, 2013.

\bibitem{raw}
M.~B. Taylor, J.~Kim, J.~Miller, D.~Wentzlaff, F.~Ghodrat, B.~Greenwald,
  H.~Hoffman, P.~Johnson, J.-W. Lee, W.~Lee, A.~Ma, A.~Saraf, M.~Seneski,
  N.~Shnidman, V.~Strumpen, M.~Frank, S.~Amarasinghe, and A.~Agarwal, ``The raw
  microprocessor: A computational fabric for software circuits and
  general-purpose programs,'' {\em IEEE Micro}, vol.~22, pp.~25--35, Mar. 2002.

\bibitem{wavescalar-sched}
M.~Mercaldi, S.~Swanson, A.~Petersen, A.~Putnam, A.~Schwerin, M.~Oskin, and
  S.~J. Eggers, ``Instruction scheduling for a tiled dataflow architecture,''
  in {\em Proceedings of the 12th international conference on Architectural
  support for programming languages and operating systems}, ASPLOS XII,
  pp.~141--150, 2006.

\bibitem{space-time-scheduler}
W.~Lee, R.~Barua, M.~Frank, D.~Srikrishna, J.~Babb, V.~Sarkar, and
  S.~Amarasinghe, ``Space-time scheduling of instruction-level parallelism on a
  raw machine,'' in {\em Proceedings of the Eighth International Conference on
  Architectural Support for Programming Languages and Operating Systems},
  ASPLOS VIII, (New York, NY, USA), pp.~46--57, ACM, 1998.

\bibitem{edge-centric-scheduler}
H.~Park, K.~Fan, S.~A. Mahlke, T.~Oh, H.~Kim, and H.-s. Kim, ``Edge-centric
  modulo scheduling for coarse-grained reconfigurable architectures,'' in {\em
  Proceedings of the 17th international conference on Parallel architectures
  and compilation techniques}, PACT '08, pp.~166--176, 2008.

\bibitem{dresc}
B.~Mei, S.~Vernalde, D.~Verkest, H.~D. Man, and R.~Lauwereins, ``Exploiting
  loop-level parallelism on coarse-grained reconfigurable architectures using
  modulo scheduling,'' {\em IEE Proceedings - Computers and Digital
  Techniques}, vol.~150, pp.~255--61--, Sept 2003.

\bibitem{ilp-scheduler}
T.~Nowatzki, M.~Sartin-Tarm, L.~De~Carli, K.~Sankaralingam, C.~Estan, and
  B.~Robatmili, ``A general constraint-centric scheduling framework for spatial
  architectures,'' in {\em Proceedings of the 34th ACM SIGPLAN Conference on
  Programming Language Design and Implementation}, PLDI '13, (New York, NY,
  USA), pp.~495--506, ACM, 2013.

\bibitem{pathfinder}
L.~McMurchie and C.~Ebeling, ``Pathfinder: A negotiation-based
  performance-driven router for fpgas,'' in {\em Third International ACM
  Symposium on Field-Programmable Gate Arrays}, pp.~111--117, Feb 1995.

\bibitem{hybrid-sched}
T.~Nowatzki, N.~Ardalani, K.~Sankaralingam, and J.~Weng, ``Hybrid
  optimization/heuristic instruction scheduling for programmable accelerator
  codesign,'' in {\em Proceedings of the 27th International Conference on
  Parallel Architectures and Compilation Techniques}, PACT '18, (New York, NY,
  USA), pp.~36:1--36:15, ACM, 2018.

\bibitem{riscv}
K.~Asanovi{\'c} and D.~A. Patterson, ``Instruction sets should be free: The
  case for risc-v,'' {\em EECS Department, University of California, Berkeley,
  Tech. Rep. UCB/EECS-2014-146}, 2014.

\bibitem{gem5}
N.~Binkert, B.~Beckmann, G.~Black, S.~K. Reinhardt, A.~Saidi, A.~Basu,
  J.~Hestness, D.~R. Hower, T.~Krishna, S.~Sardashti, R.~Sen, K.~Sewell,
  M.~Shoaib, N.~Vaish, M.~D. Hill, and D.~A. Wood, ``The gem5 simulator,'' {\em
  SIGARCH Comput. Archit. News}, 2011.

\bibitem{risc5}
A.~Roelke and M.~R. Stan, ``{RISC5}: Implementing the {RISC-V ISA} in gem5,''
  in {\em Workshop on Computer Architecture Research with RISC-V (CARRV)},
  2017.

\bibitem{trig-impl}
T.~J. Repetti, J.~a.~P. Cerqueira, M.~A. Kim, and M.~Seok, ``Pipelining a
  triggered processing element,'' in {\em Proceedings of the 50th Annual
  IEEE/ACM International Symposium on Microarchitecture}, MICRO-50 '17, (New
  York, NY, USA), pp.~96--108, ACM, 2017.

\bibitem{cacti}
R.~Balasubramonian, A.~B. Kahng, N.~Muralimanohar, A.~Shafiee, and V.~Srinivas,
  ``Cacti 7: New tools for interconnect exploration in innovative off-chip
  memories,'' {\em ACM Transactions on Architecture and Code Optimization
  (TACO)}, vol.~14, no.~2, p.~14, 2017.

\bibitem{scev}
R.~A. Van~Engelen, ``Efficient symbolic analysis for optimizing compilers,'' in
  {\em International Conference on Compiler Construction}, pp.~118--132,
  Springer, 2001.

\bibitem{energy-eff-comm}
R.~Fasthuber, F.~Catthoor, P.~Raghavan, and F.~Naessens, {\em Energy-Efficient
  Communication Processors: Design and Implementation for Emerging Wireless
  Systems}.
\newblock Springer Publishing Company, Incorporated, 2013.

\bibitem{mat-fact}
A.~Pedram, A.~Gerstlauer, and R.~van~de Geijn, ``Algorithm, architecture, and
  floating-point unit codesign of a matrix factorization accelerator,'' {\em
  IEEE Transactions on Computers}, no.~1, pp.~1--1, 2014.

\bibitem{dmt_cgra}
D.~Voitsechov and Y.~Etsion, ``Inter-thread communication in multithreaded,
  reconfigurable coarse-grain arrays,'' {\em arXiv preprint arXiv:1801.05178},
  2018.

\bibitem{sgmf}
D.~Voitsechov and Y.~Etsion, ``Single-graph multiple flows: Energy efficient
  design alternative for gpgpus,'' in {\em Proceeding of the 41st Annual
  International Symposium on Computer Architecuture}, ISCA '14, (Piscataway,
  NJ, USA), pp.~205--216, IEEE Press, 2014.

\bibitem{mt-cgrf}
D.~Voitsechov and Y.~Etsion, ``Control flow coalescing on a hybrid dataflow/von
  neumann gpgpu,'' in {\em 2015 48th Annual IEEE/ACM International Symposium on
  Microarchitecture (MICRO)}, pp.~216--227, Dec 2015.

\bibitem{pp-fpga}
R.~Prabhakar, D.~Koeplinger, K.~J. Brown, H.~Lee, C.~De~Sa, C.~Kozyrakis, and
  K.~Olukotun, ``Generating configurable hardware from parallel patterns,'' in
  {\em Proceedings of the Twenty-First International Conference on
  Architectural Support for Programming Languages and Operating Systems},
  ASPLOS '16, (New York, NY, USA), pp.~651--665, ACM, 2016.

\bibitem{swarm}
M.~C. Jeffrey, S.~Subramanian, C.~Yan, J.~Emer, and D.~Sanchez, ``A scalable
  architecture for ordered parallelism,'' in {\em 2015 48th Annual IEEE/ACM
  International Symposium on Microarchitecture (MICRO)}, pp.~228--241, Dec
  2015.

\bibitem{tao}
K.~Pingali, D.~Nguyen, M.~Kulkarni, M.~Burtscher, M.~A. Hassaan, R.~Kaleem,
  T.-H. Lee, A.~Lenharth, R.~Manevich, M.~M{\'e}ndez-Lojo, D.~Prountzos, and
  X.~Sui, ``The tao of parallelism in algorithms,'' in {\em Proceedings of the
  32Nd ACM SIGPLAN Conference on Programming Language Design and
  Implementation}, PLDI '11, (New York, NY, USA), pp.~12--25, ACM, 2011.

\bibitem{tapas}
S.~Margerm, A.~Sharifian, A.~Guha, A.~Shriraman, and G.~Pokam, ``Tapas:
  Generating parallel accelerators from parallel programs,''

\bibitem{loop-task}
J.~Kim, S.~Jiang, C.~Torng, M.~Wang, S.~Srinath, B.~Ilbeyi, K.~Al-Hawaj, and
  C.~Batten, ``Using intra-core loop-task accelerators to improve the
  productivity and performance of task-based parallel programs,'' in {\em
  Proceedings of the 50th Annual IEEE/ACM International Symposium on
  Microarchitecture}, MICRO-50 '17, (New York, NY, USA), pp.~759--773, ACM,
  2017.

\bibitem{libra}
Y.~Park, J.~J.~K. Park, H.~Park, and S.~Mahlke, ``Libra: Tailoring simd
  execution using heterogeneous hardware and dynamic configurability,'' in {\em
  Proceedings of the 2012 45th Annual IEEE/ACM International Symposium on
  Microarchitecture}, MICRO-45, (Washington, DC, USA), pp.~84--95, IEEE
  Computer Society, 2012.

\bibitem{shao}
Y.~S. Shao, S.~L. Xi, V.~Srinivasan, G.~Y. Wei, and D.~Brooks, ``Co-designing
  accelerators and soc interfaces using gem5-aladdin,'' in {\em 2016 49th
  Annual IEEE/ACM International Symposium on Microarchitecture (MICRO)},
  pp.~1--12, Oct 2016.

\end{thebibliography}

\end{document}